\documentclass[proceedings]{JHEP37}
\usepackage{eurosym}
\usepackage{amssymb}
\usepackage{amsfonts,cite}
\usepackage{amsmath}
\usepackage{graphicx}
\usepackage{multirow}
\usepackage{epsfig}
\usepackage{xcolor}

\DeclareMathOperator\sech{sech}

\newbox\mybox

\newcommand\fverb{\setbox\mybox=\hbox\bgroup\verb}
\newcommand\fverbdo{\egroup\medskip\noindent\fbox{\unhbox\mybox}\ }
\newcommand\fverbit{\egroup\item[\fbox{\unhbox\mybox}]}
\conference{Nonlinear evolution of disturbances in higher time-derivative theories}
\abstract{We investigate the evolution of localized initial value profiles when propagated in integrable versions of higher time-derivative theories. In contrast to the standard cases in nonlinear integrable systems, where these profiles evolve into a specific number of N-soliton solutions as dictated by the conservation laws, in the higher time-derivative theories the theoretical prediction is that the initial profiles can settle into either two-soliton solutions or into any number of N-soliton solutions. In the latter case this implies that the solutions exhibit oscillations that spread in time but remain finite. 
	We confirm these analytical predictions by explicitly solving the associated Cauchy problem numerically with multiple initial profiles for various higher time-derivative versions of integrable modified Korteweg-de Vries equations. In the case with the theoretical possibility of a decay into two-soliton solutions, the emergence of underlying singularities may prevent the profiles from fully developing or may be accompanied by oscillatory, chargeless standing waves at the origin.}      

\title{Nonlinear evolution of disturbances in higher time-derivative theories}
\author{Andreas Fring$^\bullet$, Takano Taira$^\circ$ and Bethan Turner$^\bullet$\\
 $\bullet$ Department of Mathematics, City, University of London, Northampton Square,\\ $\,\,$ London EC1V 0HB, UK \\
 $\circ$ Research Fellow of Japan Society for Promotion of Science, Institute of Industrial \\ $\,\,$ Science, The University of Tokyo
 5-1-5 Kashiwanoha,
  Kashiwa 277-8574, Japan\\
 
E-mail: a.fring@city.ac.uk, taira904@iis.u-tokyo.ac.jp, bethan.turner.2@city.ac.uk}

\begin{document}

\section{Introduction}

The emergence of stable soliton solutions from the evolution of generic initial profiles in continuous versions \cite{zabusky1965int} of the seminal Fermi-Pasta-Ulam-Tsingou models \cite{fermi1955studies,berman2005fermi} is one of the archetypical effects in classical nonlinear integrable field theories. The integrability of the models ensures that the system evolves into some of the $N$-soliton solutions of the underlying nonlinear integrable equation when $N>1$. Following \cite{berezin,jeffrey1972weak} one can employ the conservation laws of the model and predict how many solitons will emerge together with their amplitudes. Here, our main purpose is to investigate the analogue of this phenomenon in a set of integrable higher time-derivative theories (HTDT).

Despite the fact that HTDT unavoidably contain singularities in their classical solutions and lead to inconsistent quantum versions, they have kept being of interest because at the same time they also posses a number of very attractive features, such as being renormalizable \cite{pais1950field,stelle77ren,grav1,grav2,grav3,modesto16super}. Several proposals have been made to resolve the issues of non-normalisable states and/or the unboundedness of their spectra in the quantum version of HTDT \cite{ghostconst,salvio16quant,fakeons,bender2008no,raidal2017quantisation}. HTDT have been applied in a variety of areas in physics, such as in attempts to quantize gravity \cite{Hawking}, in applications to cosmology \cite{biswas2010towards,Salvio2,Salvio3,Salvio4}, finite temperature physics \cite{weldon98finite}, black hole solutions \cite{mignemi1992black},  BRST quantisation \cite{rivelles2003triviality,Kap1}, in a massless particle descriptions of bosons and fermions \cite{plyush89mass,Mpl} and in supersymmetric theories \cite{dine1997comments,smilga17ultrav}. Classical and quantum stability properties of HTDT were investigated in \cite{Sugg1,Sugg2,Sugg3,Sugg4,deffayet22ghost,deffayet23global}. 

There are of course many different versions of HTDT. Here we will follow a recent suggestion \cite{smilga2021exactly} and focus our investigations on a particular class of models that are obtained from exchanging space and time in Hamiltonian and higher charge systems of modified Korteweg-de Vries type. For the specific example such an idea was previously pursued in \cite{Samoilenko}. While in general the original versions of higher charge theories are still of interest in their own right \cite{bethanAF,fring23int}, we continue here our investigation from \cite{fring2024higher} on HTDT by concentrating on the study of soliton solutions in these systems 

Our manuscript is organised as follows: In section 2 we recall the standard argument of how to predict the amplitudes of the emerging solitons for a given initial value profile and discuss how this reasoning needs to be modified for HTDT. In sections 3 we carry out a detailed analytical and numerical analysis of the emergent solitons in the standard KdV system, some of its higher charge Hamiltonian systems, the integrable modified KdV system and their nonintegrable modified versions. In section 4 we carry out the adequately modified analysis for the HTDT of the systems considered in section 3. Our conclusions are stated in section 5.

\section{Emergent solitons from initial value disturbances}

We briefly recall from \cite{berezin} the standard argument of how the conservation laws of integrable systems can be used to predict the amplitudes of the emergent solitions from an initial value profile that is evolved with an integrable nonlinear equation and elaborate on how it needs to be modified for HTDT. In general, we are considering here the following Cauchy initial value problem 
\begin{equation}
\!\!\!	u_t = F(u, u_x, \ldots, u_{nx}),  \qquad u(x, t=0 )=f(x), \quad \lim_{\vert x \vert \rightarrow \infty} u(x,t), \ldots ,u_{(n-1) x }(x,t)  =0,  \label{Cauchyo}
\end{equation} 
where the function $F$ might be nonlinear in the fields $u$ and its partial $x$-derivatives up to order $n$ and the function $f(x)$ characterises the initial value profile.

The system is assumed to be integrable so that one can exploit infinitely many conservation laws of the form
\begin{equation}
   \frac{	\partial Q_\ell(x,t)}{\partial t} + \frac{\partial \chi_\ell (x,t) }{ \partial x} =0 , \qquad \ell \in \mathbb{N},
\end{equation} 
relating the charge densities $Q_\ell$ to the flux densities $ \chi_\ell $. Then ${\cal Q}_\ell(t) = \int_{- \infty}^{\infty}  Q_\ell(x,t) dx$ is conserved in time, i.e.~$d {\cal Q} /dt =0 $ for $\lim_{\vert x \vert \rightarrow \infty }  \chi_\ell (x,t)  =0$, where the latter is ensured by our asymptotic conditions in (\ref{Cauchyo}). It is well-known that any $N$-soliton solution behaves asymptotically in time as the sum of $N$ one-soliton solutions. Therefore, the corresponding charges ${\cal Q}_\ell^{(N)}(a_1,\ldots, a_N)$, depending on some parameters $a_i$, such as for instance the amplitudes, is the sum of the asymptotically acquired one-soliton contributions, i.e., ${\cal Q}_\ell^{(N)}= \sum_{i=1}^N {\cal Q}_\ell^{(1)}(a_i) $. Assuming that the initial profile breaks up into an $N$-soliton then implies that for each charge the entire initial profile charge ${\cal Q}_\ell^{(I)}$ is converted into the sum of the one-soliton contributions to that charge, i.e.,
\begin{equation}
       {\cal Q}_\ell^{(I)}=  \int_{-\infty}^{\infty} Q_\ell  \left[u(x,0) \right] dx     = \sum_{i=1}^N {\cal Q}_\ell^{(1)}(a_i) .
       \label{profevolve}
\end{equation} 
At this point it is still not determined into how many solitons $N$ the initial profile will evolve. However, each of the equations in (\ref{profevolve}) provides a constraint, which can be used to answer this question in concrete models. Moreover, one can solve the system of equations (\ref{profevolve}) for the amplitudes $a_i$ to predict them in an approximate fashion. In \cite{berezin} only combination of the lowest charges were taken into account to make theoretical predictions. However, one should stress that all possibilities need to be respected, which makes this system of equations highly overdetermined. Here we refine the analysis of \cite{berezin} by including more combinations into the analysis. Thus, our approach leads to more detailed predictions and crucially, especially for HTDT, also predicts when a break up into multi-soliton solution is prohibited.   

In HTDT the Cauchy problem (\ref{Cauchyo}) must be changed into
\begin{equation}
	0 = F(u, u_x, \ldots, u_{nx},u_t, \ldots, u_{mt}),  \qquad u(x, 0)=f_1(x), \ldots , u_{(m-1)t}(x, 0)=f_{(m-1)}(x), \label{Cauchyhtdt}
\end{equation} 
together with adequate boundary conditions. All $m$ functions $f_i$ are independent. Following \cite{Smilga6,Smilgaacta,fring2024higher} we assume here in the first instance that the HTDT is obtained from (\ref{Cauchyo}) by exchanging time and space, i.e. $x \leftrightarrow t$. This approach allows us to use the conservation laws (\ref{profevolve}) with $ Q_\ell(x,t) \leftrightarrow   \chi_\ell(t,x)  $ and make similar prediction for the number of solitons and their amplitudes in these theories into which the initial profiles $f_1, \ldots f_m$ evolve. The interesting aspect to be investigated here is how the different types of singularities, that are inevitable present in a HTDT, manifest themselves in this break up process.   

\section{Emergent solitons in modified Korteweg-de Vries systems}
We start our investigation with the series of the modified KdV system in the form
\begin{equation}
	u_t + n(n-1) u^{n-2} u_x + u_{xxx} =0 , \qquad n \in \mathbb{N}.  \label{stanKdV}
\end{equation}  
Rescaling equation (\ref{stanKdV}) by 
\begin{equation}
	x \rightarrow \frac{\sigma ^2 	\lambda_n ^{3-2 n}}{(n-1)^2
		n^2} x, \quad 
	t \rightarrow 	\lambda_n t, \quad
	u \rightarrow \lambda_n    u  , \quad \lambda_n := 	\left[ \frac{(n-1)^3 n^3}{\sigma ^2}\right]^{\frac{1}{4-3 n}} , \label{scaling}
\end{equation} 
we obtain 
\begin{equation}
	u_t +  u^{n-2} u_x + \frac{1}{\sigma^2} u_{xxx} =0 .   \label{stanKdVtr}
\end{equation} 
Next we solve the Cauchy problem for equation (\ref{stanKdVtr}) with initial value profile $u(x, t=0)=f(x)$ and vanishing asymptotic values $\lim_{\vert x \vert \rightarrow \infty} u(x,t), u_t(x,t) , u_{tt}(x,t)  =0 $. In accordance with the similarity principle the parameter $\sigma$, that was introduced through the scaling (\ref{scaling}), is known to separate regions of different characteristic behaviour \cite{berezin}. Letting the initial profile evolve by means of (\ref{stanKdVtr}), the integrability of the models for $n=3,4$ ensures that the profile will eventually settle into a multi-soliton solution and hence for large times into a number of one-soliton solutions. For the nonintegrable systems with $n>4$ no solitons are expected to emerge. 

\subsection{Emergent solitons in the Korteweg-de Vries system}
\begin{figure}[h]
 	\centering         
 	\begin{minipage}[b]{0.9\textwidth}      
 		\includegraphics[width=\textwidth, height=8cm]{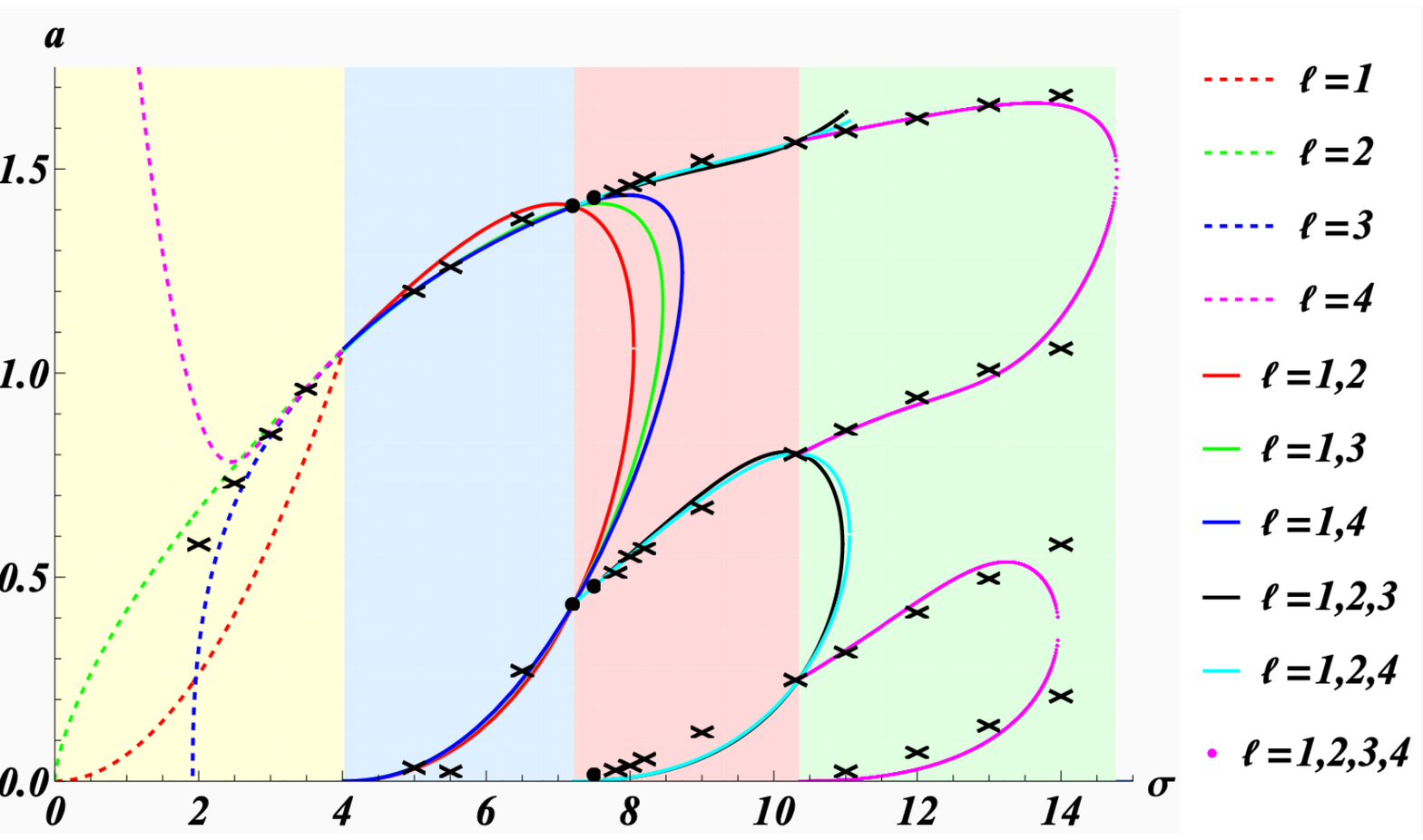}
 	\end{minipage}   
 	\caption{Domains of $N$-soliton states emerging from an initial Gaussian profile in the KdV system together with their predicted amplitudes. The nonsoliton, two-soliton, three-soliton and four-soliton regions are shaded in yellow, blue, red and green, respectively. The black crosses and dots represent the values of the amplitudes from the actual numerical solutions of the Cauchy problem for specific values of $\sigma$. For the values of the black dots we show the explicit solutions in figure \ref{sig772}. } 
 	\label{solbounds}
 \end{figure}

The first case we consider is to revisit the standard KdV-equation corresponding to the equation of motion (\ref{stanKdV}) with $n=3$. We recall from \cite{whitham1965non,miura1968kortxx,miura1968korteweg} the charge and flux densities of the first four conserved quantities, which when appropriately scaled become 
\begin{eqnarray}
	 Q_1&=& u, \qquad \qquad  \qquad \qquad  \qquad  \qquad \,\,\,	\chi_1= \frac{1}{2} u^2 +  \frac{1}{\sigma^2}   u_{xx}, \label{KdVcharges} \\
	 Q_2 &=& \frac{1}{2} u^2, \qquad \qquad   \qquad\qquad \qquad \quad \,\,\, \chi_2= \frac{1}{3} u^3 +  \frac{1}{2 \sigma^2}   \left( 2 u u_{xx} -   u_{xx}^2   \right), \notag \\
	 		Q_3&=&\frac{1}{3} u^3 - \frac{1}{\sigma^2} u_x^2, \qquad \qquad  \qquad \qquad
	 		\chi_3=  \frac{1}{4} u^4  +  \frac{1}{\sigma^2}   \left( u^2 u_{xx}  + 2 u u_x    \right)  +   \frac{1}{\sigma^4}  u_{xx}^2, \notag \\
		Q_4&=&\frac{1}{4} u^4 - \frac{3}{\sigma^2} u u_x^2 +  \frac{9}{ 5 \sigma^4}  u_{xx}^2, \quad \notag \\
		\chi_4 &=& \frac{1}{5} u^5 + \frac{1}{\sigma^2} \left( u^3 u_{xx} -\frac{9}{2} u^2 u_x^2   \right)  + \frac{3}{\sigma^4}  \left(  u_x^2 u_{xx}-2 u u_x u_{xxx}+\frac{8 u u_{xx}^2}{5}  \right)  \qquad   \qquad \notag \\
		&&	-\frac{9}{ 5 \sigma^6}  \left(u_{xxx}^2-2 u_{xx} u_{xxxx}\right) .\notag
\end{eqnarray} 
For the one-soliton solution of (\ref{stanKdVtr})
\begin{equation}
       u(x,t) = a \sech^2\left[ \frac{\sigma}{\sigma_s}  \sqrt{a} \left(  x - \frac{a}{3} t     \right)      \right],  \label{solonestand}
\end{equation} 
with nonlinearity index $\sigma_s = \sqrt{12}$, we compute with (\ref{KdVcharges}) the conserved charges
\begin{equation}
        {\cal Q}_1 =  \frac{2 \sigma_s \sqrt{a}}{\sigma }  , \qquad 
        {\cal Q}_2 = \frac{2 \sigma_s a^{3/2}}{3 \sigma }, \qquad 
         {\cal Q}_3 = \frac{4 \sigma_s a^{5/2}}{15\sigma },  \qquad 
         {\cal Q}_4 = \frac{4 \sigma_s a^{7/2}}{35 \sigma }. \label{solcharge}
\end{equation} 
Thus, if the initial profile would be converted entirely into a one-soliton solution, relation (\ref{profevolve})  implies that $  {\cal Q}_\ell = {\cal Q}_\ell^{(I)} $. In principle, these relations could be used to predict the amplitude of the emerging soliton. However, when solved for the amplitudes as functions of $\sigma$ these equations lead to vastly mismatching solutions for different values of $\ell$ and hence the solution is not unique, see the yellow region in figure \ref{solbounds} for some examples. The marked amplitudes of some solitary waves obtained from the numerical solutions are very crudely identified, as we have ignored the typical oscillatory tail that spreads to negative infinity as time evolves. This means the various constraints imposed by the integrability of the system prevent a full conversion of the initial profile into a one-soliton, so that the region $\sigma < \sigma_c$ in which no multi-soliton can form, is referred to as a ``nonsoliton'' region \cite{berezin}. In this region the initial disturbance decays into an oscillating wave that spreads throughout space.

Assuming instead that the initial profile evolves into a two-soliton with amplitudes $a_1$ and $a_2$, relation (\ref{profevolve}) yields the two constraining equations from the first two equation with $\ell=1$ and $\ell=2$
\begin{equation}
	 \sqrt{a_1} + \sqrt{a_2} =  \frac{\sigma}{ 2 \sigma_s} {\cal Q}_1^{(I)} =: I_1, \qquad \text{and} \qquad 
	 a_1^{3/2} +  a_2^{3/2} =   \frac{3 \sigma}{ 2 \sigma_s} {\cal Q}_2^{(I)}  =: I_2,     \label{x45}
\end{equation} 
which are easily solved to
\begin{equation}
a_{1/2}  = \left( \frac{ I_1 }{2} \pm  \frac{\sqrt{ 4   I_1   I_2 - I_1^4  }}{2 \sqrt{3} I_1}  \right)^2.
\end{equation}
Demanding the amplitudes to be real and its square roots to be positive, as assumed in (\ref{x45}), gives the following interval for $\sigma$ in which the initial profile may consistently evolves into two-soliton solutions
\begin{equation}
\sigma_c <   \sigma  <   2 \sigma_c , \qquad \sigma_c = 12  \sqrt{  {\cal Q}_2^{(I)}/ \left({\cal Q}_1^{(I)} \right)^3} . \label{twosolreg}
\end{equation} 
For a Gaussian initial profile $f(x) = e^{-x^2}$ we obtain from (\ref{profevolve})
\begin{equation}
	 {\cal Q}_1^{(I)} = \sqrt{\pi}, \quad {\cal Q}_2^{(I)}  =   \sqrt{ \frac{\pi}{2^{3}}  },      \quad {\cal Q}_3^{(I)}  = \sqrt{ \frac{\pi}{3^{3}}  } -\frac{1}{\sigma^2} \sqrt{\frac{\pi}{2}}, \quad
	{\cal Q}_4^{(I)}  = \frac{\sqrt{\pi}}{8}-\frac{2}{\sigma^2} \sqrt{\frac{\pi}{3}}+\frac{1}{\sigma^4} \frac{27}{5} \sqrt{\frac{\pi}{2}}.
\end{equation}
 This information is sufficient to predict the amplitudes in the different $N$-soliton regions.  
 
 The bounds for the two-soliton region (\ref{twosolreg}) are then characterised by $ \sigma_c= 6 \times 2^{1/4}/\sqrt{\pi} \approx 4.026 $, which is in agreement with \cite{berezin}. However, here we refine this analysis and consider also solutions from combining conservation laws for different values of $\ell$. In figure \ref{solbounds} we have included for a variety of combinations the numerically obtained predicted amplitudes together with the actual numerical solutions of the initial value problem. 
 
\begin{figure}[h]
 	\centering         
 	\begin{minipage}[b]{0.49\textwidth}      
 		\includegraphics[width=\textwidth]{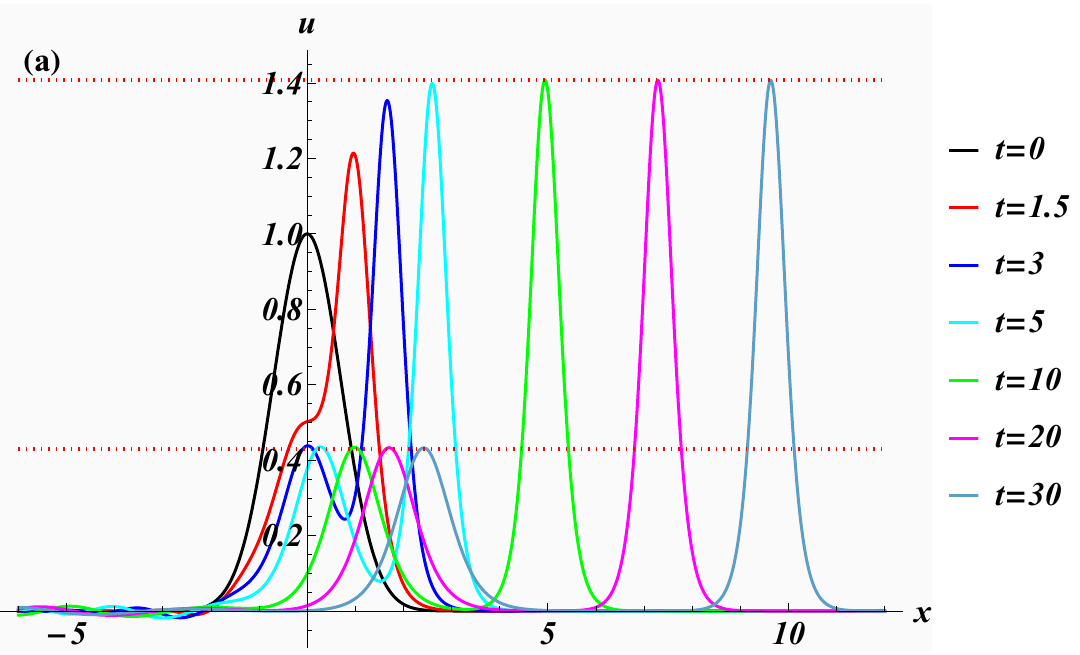}
 	\end{minipage}   
 	\begin{minipage}[b]{0.49\textwidth}      
 		\includegraphics[width=\textwidth]{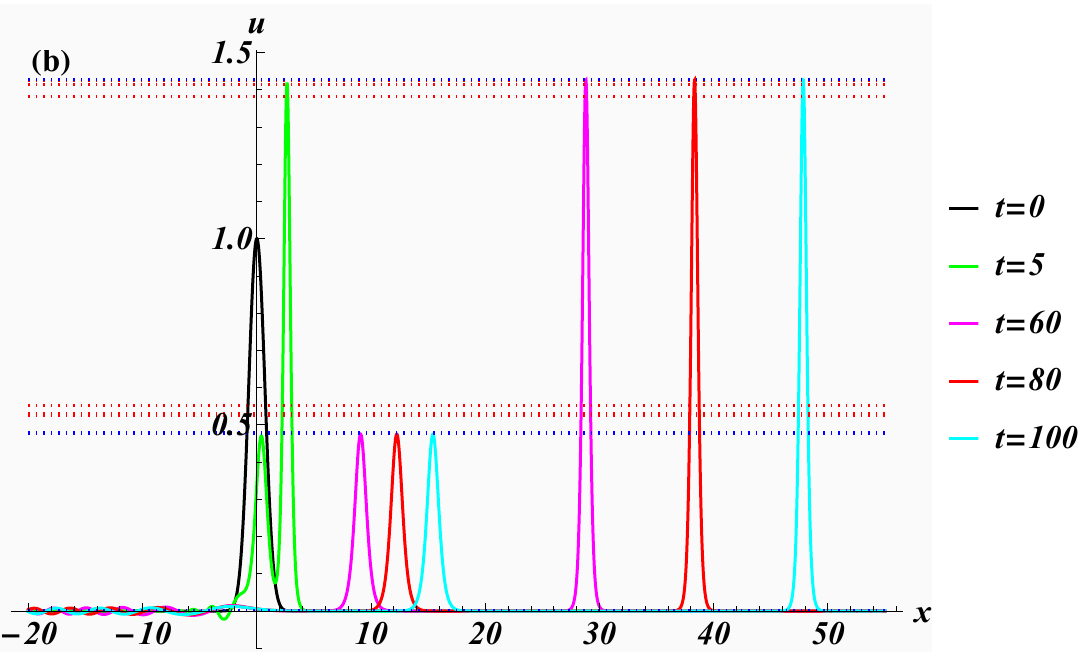}
 	\end{minipage}   
 	\caption{Evolution of an initial Gaussian profile into a two-soliton solution panel (a) and three-solition solution panel (b) as time evolves for the KdV system for $\sigma=7.2$ and $\sigma=7.5$, respectively. The predicted two and three-soliton amplitudes are depicted as dotted red and blue lines, respectively.} 
 	\label{sig772}
 \end{figure}

 For the two-soliton case we observe that, unlike as in the nonsoliton region, there are regions for which the predicted amplitudes from different combinations roughly coincide. For $\sigma_c< \sigma \lessapprox 7.22$ the two one-soliton solutions are formed with small deviations from the anticipated amplitudes because each combination of the conservation laws leads to slightly different predictions. For $\sigma \approx 7.22$ the agreement is extremely good, since all the predicted amplitudes almost exactly coincide, the system is left with no ambiguities into which solution to settle, see figure \ref{sig772} panel (a). 
 
 However, in the region $\sigma\gtrsim 7.22$ the predictions start to differ more significantly. Moreover, beyond that value even three-soliton solution may occur, hence the ``three-soliton region" is partially encroaching into the ``two-soliton region" that was predicted in  \cite{berezin}. In figure \ref{solbounds} we have also included some solutions for these cases  computed numerically from solving the three equations
\begin{equation}
\sum_{i=1}^3 \sqrt{a_i} =  \frac{\sigma}{ 2 \sigma_s} {\cal Q}_1^{(I)},\qquad
\sum_{i=1}^3 a_i^{3/2}  =   \frac{3 \sigma}{ 2 \sigma_s} {\cal Q}_2^{(I)},  \qquad
	\sum_{i=1}^3 a_i^{5/2} =   \frac{15 \sigma}{ 4 \sigma_s} {\cal Q}_3^{(I)} . 
\end{equation} 
We see in figure \ref{solbounds} that the predicted amplitudes from combining different combinations of the conservation laws matches quite well the actual numerical solution. We also note in panel (b) of figure \ref{sig772} that the acquired values in the ``two-soliton region" are in fact those predicted for the three-soliton with one of the amplitudes being very small so that the solutions only appears to be a two-soliton. We have also included the predictions of the four-soliton solutions which start to emerge at around $\sigma \approx 10.355$ where all the three-soliton predictions and the largest amplitudes of the four-soliton prediction coincide.

The observed features suggest more generally that {\em an initial profile will always break up into the maximal number of one-solitons that is allowed by the conservation laws (\ref{profevolve}).}

\subsection{Emergent solitons in higher charge KdV Hamiltonian systems}
Next we interpret the higher KdV charge ${\cal Q}_4$ as a Hamiltonian. In order to derive Hamilton's equation of motion we need to identify the canonical fields. Here we may achieve this by a direct extrapolation from the standard Hamiltonian system \cite{Nutku}, for a more general treatment see \cite{fring2024higher}. Multiplying this charge by $-1/3$, introducing the canonical momentum field $\pi = \psi_x/2$ by adding zero to it and replacing $u \rightarrow \psi_x$, we obtain the higher charge Hamiltonian density
\begin{equation}
	{\cal H}_4 = \pi \psi_t- \frac{1}{2} \psi_t \psi_x- \frac{1}{12} \psi_x^4 + \frac{1}{\sigma^2}  \psi_x \psi_{xx}^2 - \frac{3}{5 \sigma^4} \psi_{xxx}^2 .
\end{equation}
The corresponding Hamilton's equations resulting from this Hamiltonian are
\begin{eqnarray}
\psi_t  &=& \frac{\delta {\cal H}_4}{\delta \pi}  =\frac{\partial {\cal H}_4  }{\partial \pi} = \psi_t , \\
\pi_t  &=& -\frac{\delta {\cal H}_4 }{\delta \phi}=- \left[  \frac{\partial  {\cal H}_4  }{ \partial \psi}  -   \partial_x \left(  \frac{\partial  {\cal H}_4  }{ \partial \psi_x}   \right) +   \partial_{x}^2 \left(  \frac{\partial  {\cal H}_4  }{ \partial \psi_{xx} }   \right) -   \partial_{x}^3 \left(  \frac{\partial  {\cal H}_4  }{ \partial \psi_{xxx} }   \right)  \right] \\
&=&  -\frac{1}{2} \psi_{xt} - \frac{1}{3} (\psi_x^3)_x  + \frac{1}{\sigma^2}  (\psi_{xx}^2 )_x - \frac{2}{\sigma^2} (\psi_x \psi_{xx})_{xx} - \frac{6}{5 \sigma^4} (\psi_{xxx} )_{xxx} . \notag
\end{eqnarray}
In terms of the standard field $u$ the equation of motion reads
\begin{equation}
u_t + u^2 u_x +\frac{2}{ \sigma^2} \left[ (u_x^2)_x  + u u_{xxx}   \right]  + \frac{5}{6 \sigma^4} u_{5x} =0 . \label{higherQKdV}
\end{equation}
We find a one-soliton solution for (\ref{higherQKdV})
\begin{equation}
	u(x,t) = a \sech^2\left[ \frac{\sigma}{\sigma_s}  \sqrt{a} \left(  x - \frac{2 a^2}{15} t     \right)      \right].  \label{hisol}
\end{equation} 
The charges obtained from integrating the densities (\ref{KdVcharges}) are also conserved, subject to the equation of motion (\ref{higherQKdV}). The fluxes will of course change. We report the first three expressions
\begin{eqnarray}
\chi_1 &=& \frac{u^3}{3}+ \frac{1}{\sigma^2} \left( u_x^2 + 2 u u_{xx} \right)    + \frac{6 }{5 \sigma ^4}  u_{4 x} ,\\
\chi_2 &=&  \frac{u^4}{4}+\frac{2}{\sigma^2}   u^2 u_{xx}+ \frac{3}{ 5 \sigma^4}  \left( 2u u_{4 x}- 2 u_x u_{xxx} + u_{xx}^2 \right) ,  \\
\chi_3 &=&  \frac{u^5}{5}+ \frac{2}{\sigma^2}   u^2 \left(u u_{xx}-u_x^2\right)
+   \frac{2 }{5 \sigma^4}    \left[ u
	\left(3 u u_{4 x}+8 u_{xx}^2\right)-14 u_x^2 u_{xx}-16 u u_x u_{xxx}\right]  \qquad \\
	&&
	+\frac{6 }{5 \sigma ^6}   \left(2 u_{4 x}
		u_{xx}-2 u_x u_{5 x}-u_{xxx}^2\right)  . \notag
\end{eqnarray}
Since the static part of the two one-soliton solutions (\ref{solonestand}) and (\ref{hisol}) coincide, and also the general expression for the corresponding charges are identical, the predictions for the amplitudes and the number of one-solitons to emerge are the same. The only difference between (\ref{solonestand}) and (\ref{hisol}) are the soliton speeds $v_1=a/3$ and $v_2=2a^2/15$, respectively. Since $v_2 <v_1$ for $0<a<5/2$ and the upper bound for any of the acquired one-soliton amplitudes is 2, see \cite{karpman67asymp}, the solitons in the higher charge Hamiltonian system will always be identical in height but slower than those in the original KdV equation. We have verified this numerically.

\subsection{Emergent solitons in the modified Korteweg-de Vries equation}
\begin{figure}[h]
	\centering         
	\begin{minipage}[b]{0.95\textwidth}      
		\includegraphics[width=\textwidth, height=8cm]{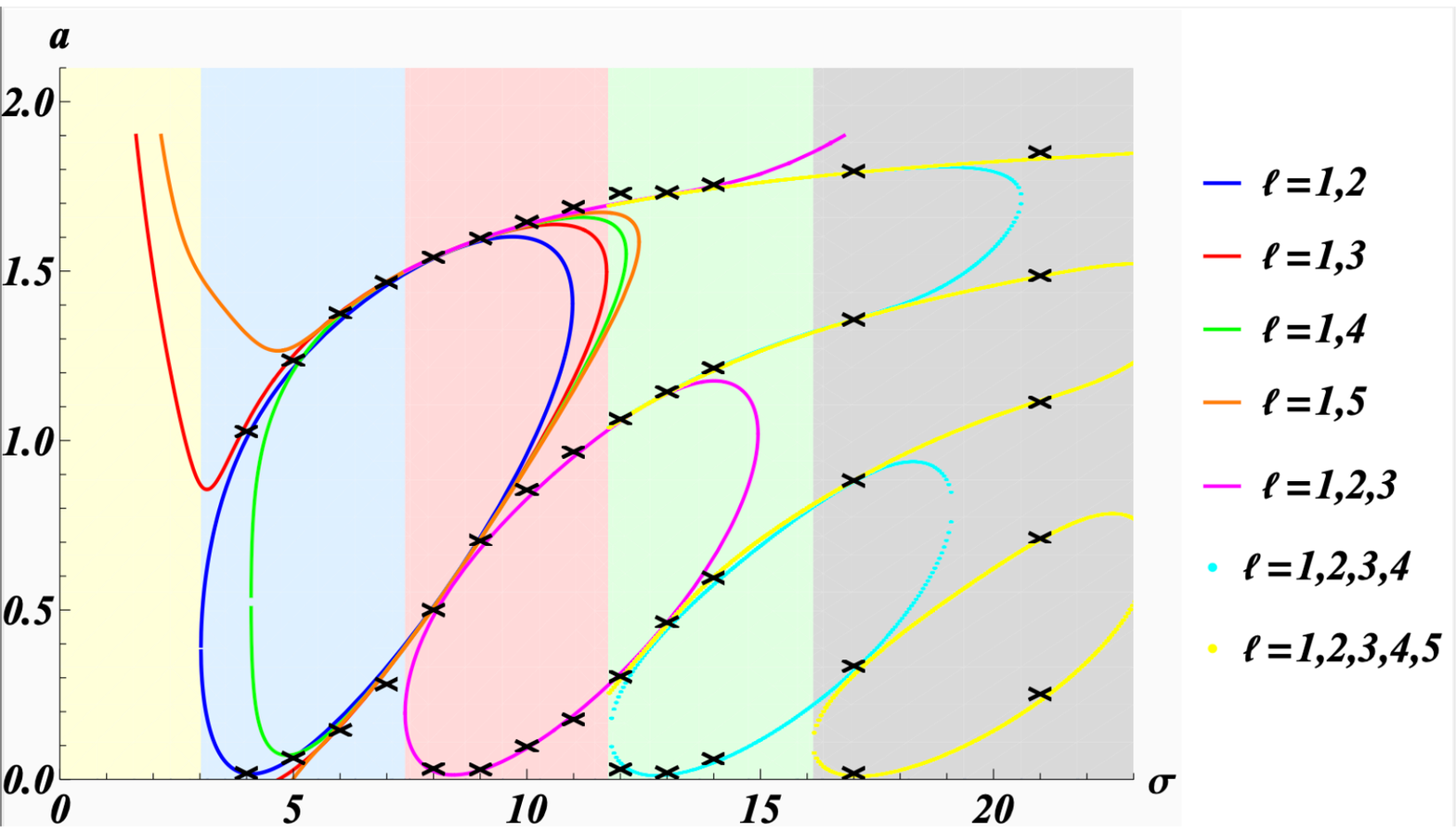}
	\end{minipage}   
	\caption{Domains of $N$-soliton states emerging from an initial Gaussian profile in the mKdV system together with their predicted amplitudes. Colour conventions are the same as in figure \ref{solbounds} with the addition of the five-soliton region in grey.}
	\label{solboundsmK}
\end{figure}

Next we consider modified KdV-equation corresponding to (\ref{stanKdV}) with $n=4$, which in standard terminology is the original modified KdV equation.
The associated charge and flux densities of the first five conserved quantities can be found in \cite{whitham1965non,miura1968kortxx,miura1968korteweg}, and when scaled appropriately read 
\begin{eqnarray}
	Q_1&=& \frac{1}{2} u^2, \qquad  \qquad  \qquad 	\chi_1= \frac{1}{4} u^4 +  \frac{1}{\sigma^2}  \left(  u u_{xx}  - \frac{1}{2} u_x^2   \right)    , \\
	Q_2 &=& \frac{1}{4} u^4 - \frac{3}{ 2 \sigma^2 } u_x^2, \qquad  \chi_2= \frac{1}{6}  u^6 + \frac{1}{\sigma^2} \left( u^3 u_{xx}-3 u^2 u_x^2 \right)  
	 +  \frac{3}{\sigma^4} \left(  \frac{1}{2} u_{xx}^2 - u_x u_{xxx}    \right), \notag \\
	Q_3&= &  \frac{1}{6} u^6 -\frac{5 }{\sigma ^2} u^2 u_x^2 +\frac{3 }{\sigma ^4} u_{xx}^2 , \notag \\
	\chi_3&=& \frac{1}{8} u^8
	+ \frac{1}{2 \sigma^2}  \left( 2 u^5 u_{{xx}}-15 u^4 u_x^2 \right)
	+  \frac{1}{2 \sigma^4}   \left[ 4 u^2 \left(4 u_{{xx}}^2-5 u_x u_{{xxx}}\right)
		+20 u u_x^2 u_{{xx}}+ u_x^4 \right] \notag	\\
		&&+ \frac{1}{\sigma^6} \left( 6 u_{{xx}}	u_{{xxxx}}-4 u_{{xxx}}^2   \right), \notag \\
	Q_4	&=& \frac{u^8}{8}-\frac{21 }{2 \sigma ^2}u^4 u_x^2 -\frac{63}{10 \sigma
			^4}  \left(u_x^4-2 u^2 u_{xx}^2\right) -\frac{27 }{5 \sigma ^6} u_{xxx}^2 ,\notag \\
	\chi_4&=& \frac{1}{10} u^{10}+  \frac{1}{\sigma ^2} u^6 \left(u u_{xx}-14 u_x^2\right)
		-\frac{21 }{10 \sigma ^4} u^2 \left(10 u^2 u_x u_{xxx}-11 u^2
		u_{xx}^2-20 u u_x^2 u_{xx}+12 u_x^4\right)\notag \\
		&& 	-\frac{9 }{5 \sigma ^6} \left(10 u^2 u_{xxx}^2-14 u^2 u_{4 x}
		u_{xx}+28 u u_x u_{xx} u_{xxx}+u_x^2 u_{xx}^2+14 u_x^3
		u_{xxx}-2 u u_{xx}^3\right) \notag \\
		  &&	+\frac{27 }{5 \sigma ^8} \left(u_{4 x}^2-2 u_{5 x} 
		  u_{xxx}\right) \notag , \\
		  {\cal Q}_5 &=&   \frac{1}{10} u^{10} -\frac{18 }{\sigma ^2} u^6 u_x^2 
		        +\frac{18 }{5
		  	\sigma ^4}  u^2 \left(9 u^2 u_{xx}^2-19 u_x^4\right)  
		  	   +\frac{108 }{35
		  	\sigma ^6} \left(51 u_x^2 u_{xx}^2+20 u u_{xx}^3  -9 u^2 u_{xxx}^2  \right)    \notag  \\
		  	&& +\frac{324 }{35 \sigma ^8}   u_{4 x}^2   .  \notag
\end{eqnarray} 
We will not report the flux $\chi_5$ here as it is rather lengthy. For the one-soliton solution of (\ref{stanKdV}) with $n=4$
\begin{equation}
	u(x,t) = a \sech\left[ \frac{a \,\sigma}{ \sqrt{6}} \left(  x - \frac{a^2}{6} t     \right)      \right],
\end{equation} 
we compute the values of the conserved charges
\begin{equation}
	{\cal Q}_1 =  \frac{\sqrt{6} \,a }{\sigma }  , \qquad 
	{\cal Q}_2 = \frac{a^{3}}{\sqrt{6} \sigma }, \qquad 
	{\cal Q}_3 = \frac{a^{5}}{5 \sqrt{6}  \sigma }, \quad 
	{\cal Q}_4 = \frac{ \sqrt{3} a^{7}}{70 \sqrt{2}  \sigma },  \quad
		{\cal Q}_5 = \frac{ a^{9}}{105 \sqrt{6}  \sigma },  \label{concharmKdV}
\end{equation} 
and also the values the charges (\ref{profevolve}) acquire with the same Gaussian initial profile as previously
\begin{eqnarray}
	{\cal Q}_1^{(I)} &=& \frac{1}{2} \sqrt{\frac{\pi }{2}} , \qquad 
	{\cal Q}_2^{(I)} = \frac{\sqrt{\pi }}{8}-\frac{3 }{2 \sigma ^2} \sqrt{\frac{\pi }{2}} , \qquad 
	{\cal Q}_3^{(I)} = \frac{1}{6}\sqrt{\frac{\pi }{6}} -\frac{5 \sqrt{\pi }}{4 \sigma ^2}+ \frac{9 }{\sigma ^4} \sqrt{\frac{\pi }{2}} ,  \qquad \\
  {\cal Q}_4^{(I)} &=& \frac{1}{16} \sqrt{\frac{\pi }{2}}-\frac{7 }{2 \sigma ^2} \sqrt{\frac{\pi
  	}{6}} + \frac{1197 \sqrt{\pi }}{80 \sigma ^4}	  -\frac{81 }{\sigma ^6} \sqrt{\frac{\pi }{2}} . \notag \\
  {\cal Q}_5^{(I)} &=&\frac{1}{10} \sqrt{\frac{\pi }{10}} -\frac{9 }{4 \sigma ^2}  \sqrt{\frac{\pi }{2}} +\frac{62 \sqrt{6 \pi }}{5 \sigma
  	^4}  -\frac{15741 \sqrt{\pi }}{70 \sigma ^6}+ \frac{486 \sqrt{2 \pi } }{\sigma ^8}
\end{eqnarray} 
As explained in much detail in the previous sections, we use the charge conservation equation (\ref{profevolve}) for various combinations to determine the different soliton regions. Taking $\ell=1,2$ we compute the predicted two-soliton amplitudes to
\begin{equation}
	a_\pm = \frac{\sqrt{3 \pi } \sigma ^2 \pm \sqrt{ 288 \sqrt{2} \sigma ^2  -\pi  \sigma ^4-3456}}{24 \sigma } ,
\end{equation}
For $ 3.0213 \approx  \left( 24 \sqrt{2} \left(6-\sqrt{36-3 \pi }\right) / \pi   \right)^{1/2} \leq  \sigma \leq  \left( 24 \sqrt{2} \left(6+\sqrt{36+3 \pi }\right) / \pi   \right)^{1/2} \approx  10.9781 $ these amplitudes are real. We also compute the predicted three-soliton amplitudes by taking $\ell=1,2,3$, which turn out to be real for $ 7.392064   \leq \sigma \leq  14.947399 $. Thus compared to the prediction for the KdV equation we have a much larger overlap between the two and three-soliton region. A comparison with the actual numerical results from evolving the initial profile is presented in figure \ref{solboundsmK}. 

We observed that once a three-soliton is possible to occur, the numerical solutions do in fact settle into them. The same holds for the predicted four-soliton amplitude predictions, that are also included into figure  \ref{solboundsmK}, when compared to the three soliton predictions. This observation confirms the general statement made at the end of section 3.1, that an initial profile always breaks up into the maximal possible number of multi-solitons.

\subsection{Solitary waves in the nonintegrable modified KdV equations}
\begin{figure}[h]
	\centering         
	\begin{minipage}[b]{0.49\textwidth}     
		\includegraphics[width=\textwidth ]{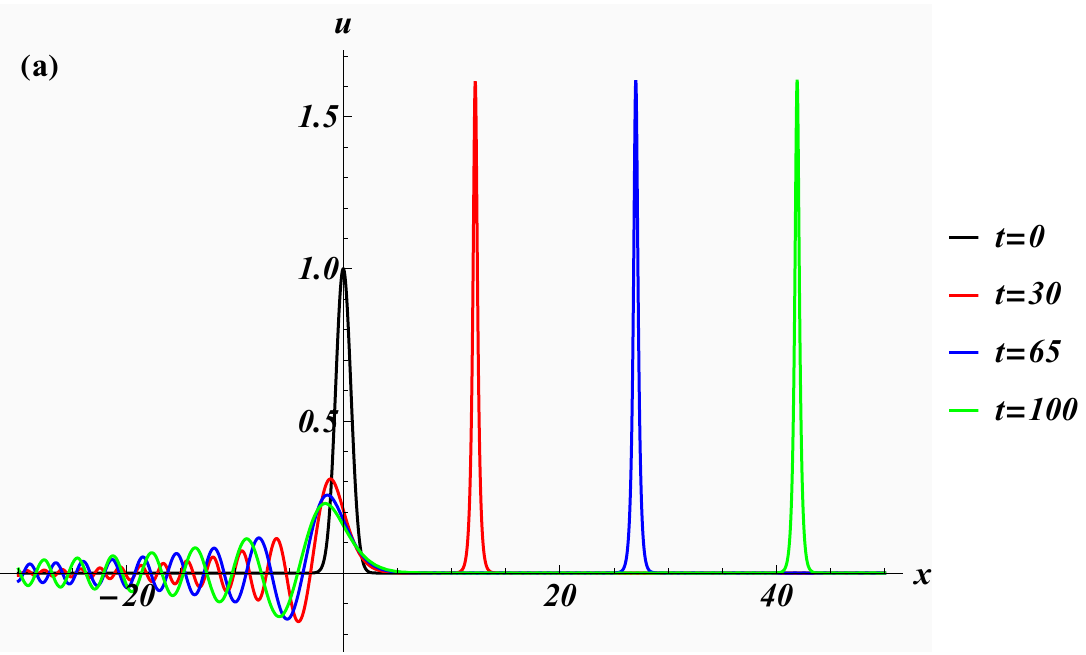}
	\end{minipage}   
	\begin{minipage}[b]{0.49\textwidth}     
		\includegraphics[width=\textwidth]{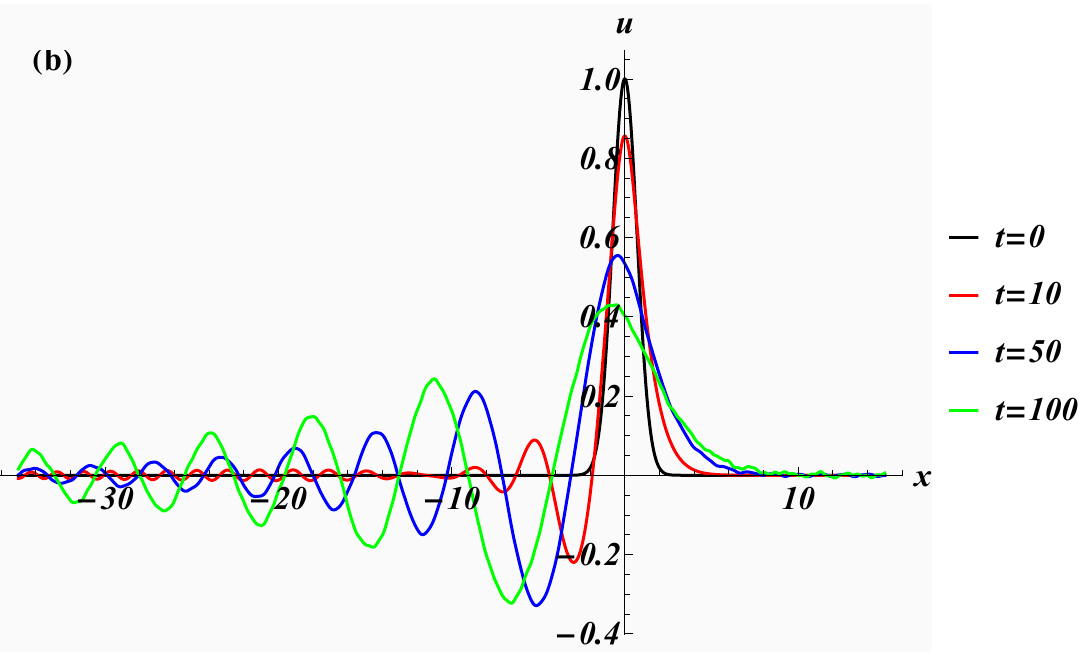}
	\end{minipage}   
	\caption{Evolution of an initial Gaussian profile in nonintegrable versions of the modified KdV equations with $n=5$, $\sigma =7$ and $n=6$, $\sigma =4$ in panel (a) and panel (b), respectively.}
	\label{nonint34}
\end{figure}

For completeness we also present here two examples for a nonintegrable version of (\ref{stanKdV}), i.e. for $n>4$. For these values the initial profile does not break up into a multi-soliton solution. However, the characteristic behaviour for $n=5$ is different from the other cases
as exemplified for two cases in figure \ref{nonint34}. In the $n=5$ case, panel (a), the solution behaves very much like the integrable cases in the nonsoliton region, i.e. the initial disturbance settles into a moving solitary wave, but also maintains an oscillatory tail for negative $x$. However, even for larger $\sigma$ we did not observe any break up into multi-soliton solutions, which is of course a signature of the model not being integrable.  In contrast, in the other cases the initial disturbance only transforms into an oscillatory tail that stretches more and more in space as time evolves. No solitary waves are emerging in these cases, see panel (b) for an example. 

\section{Emergent solitons in HTDT versions of modified KdV equations}
Next we consider the rotated version of equation (\ref{stanKdVtr}) with time and space exchanged
\begin{equation}
	u_x +  u^{n-2} u_t + \frac{1}{\sigma^2} u_{ttt} =0 .   \label{rotstanKdVtr}
\end{equation} 
and solve the rotated Cauchy problem (\ref{Cauchyhtdt}) for this equation with initial value profile $u(x, t=0)=f_1(x)$, $u_t(x, t=0)=f_2(x)$, $u_{tt}(x, t=0)=f_2(x)$,  and vanishing asymptotic values $\lim_{\vert x \vert \rightarrow \infty} u(x,t)  =0 $. Equivalently, we may of course also rotate (\ref{stanKdV}) and change the scaling (\ref{scaling}) appropriately to obtain (\ref{rotstanKdVtr}).

\subsection{Emergent solitons in the HTDT version of the KdV system}
 As mentioned in section 2, in the rotated case the conserved quantities are the same as in the original equation with $x \leftrightarrow t$ and $Q_\ell (x,t) \leftrightarrow \chi_\ell (t,x)$, i.e., the first  charge and flux densities in the rotated case for $n=3$ read
 \begin{equation}
 	Q_1= \frac{1}{2} u^2 +  \frac{1}{\sigma^2}   u_{tt}, \,\,	\chi_1= u,  \quad
 	Q_2= \frac{1}{3} u^3 +  \frac{1}{2 \sigma^2}   \left( 2 u u_{tt} -   u_{tt}^2   \right),  \,\, 	\chi_2 = \frac{1}{2} u^2, \quad \text{etc.}
 \end{equation}
We find the following one-soliton solution to (\ref{rotstanKdVtr}) for $n=3$
\begin{equation}
	u(x,t) = a \sech^2\left[ \frac{\sigma}{\sigma_s} \frac{a^{3/2}}{3}   \left(  x- \frac{3}{a}  t     \right)     \right].  \label{rotsol}
\end{equation} 
In general, the solutions for the rotated KdV system were found to be unstable \cite{Smilga6,Smilgaacta}, in the sense that they develop singularities of different type, as was discussed in detail in \cite{fring2024higher} for the periodic solution in terms of Jacobi elliptic functions. Here we investigate how these features manifest themselves for the emergent soliton solution. At first we solve the rotated Cauchy problem by implementing the profiles directly from the exact solution (\ref{rotsol})
\begin{eqnarray}
	u(x,0)&=&a \, \text{sech}^2\left(\frac{a^{3/2} \sigma  x}{6 \sqrt{3}}\right), \quad   u_t(x,0)=\frac{a^{3/2} \sigma }{\sqrt{3}} \tanh \left(\frac{a^{3/2} \sigma  x}{6 \sqrt{3}}\right) \text{sech}^2\left(\frac{a^{3/2}
			\sigma  x}{6 \sqrt{3}}\right), \,\,\,\,\,  \,\,\,\,\,   \label{rotexcauch3} \\
		  u_{tt}(x,0)&=&\frac{1}{6} a^2 \sigma ^2 \left[ \cosh \left(\frac{a^{3/2} \sigma  x}{3 \sqrt{3}}\right)-2\right]
		  \text{sech}^4\left(\frac{a^{3/2} \sigma  x}{6 \sqrt{3}}\right), \,\,\, \quad \lim_{\vert x \vert \rightarrow \infty}u( x,t) = 0. \notag
\end{eqnarray}
Since the one-soliton solutions have finite compact support, the latter boundary value can be implemented numerically to a very high precision simply by taking the finite values of the interval in $x$ to be very large. Thus, unlike as for periodic solutions of elliptic type, for the one-soliton solution the initial boundary value problem becomes a genuine Cauchy problem even when tackled numerically. 

\begin{figure}[h]
	\centering         
	\begin{minipage}[b]{0.49\textwidth}      
		\includegraphics[width=\textwidth]{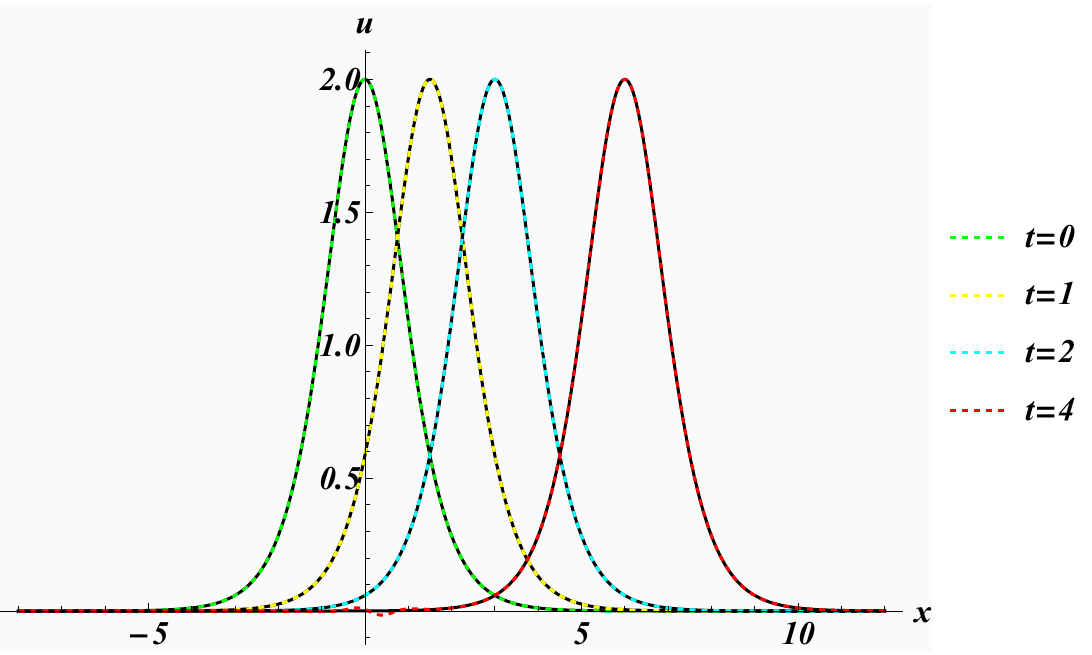}
	\end{minipage}   
	\begin{minipage}[b]{0.49\textwidth}      
		\includegraphics[width=\textwidth]{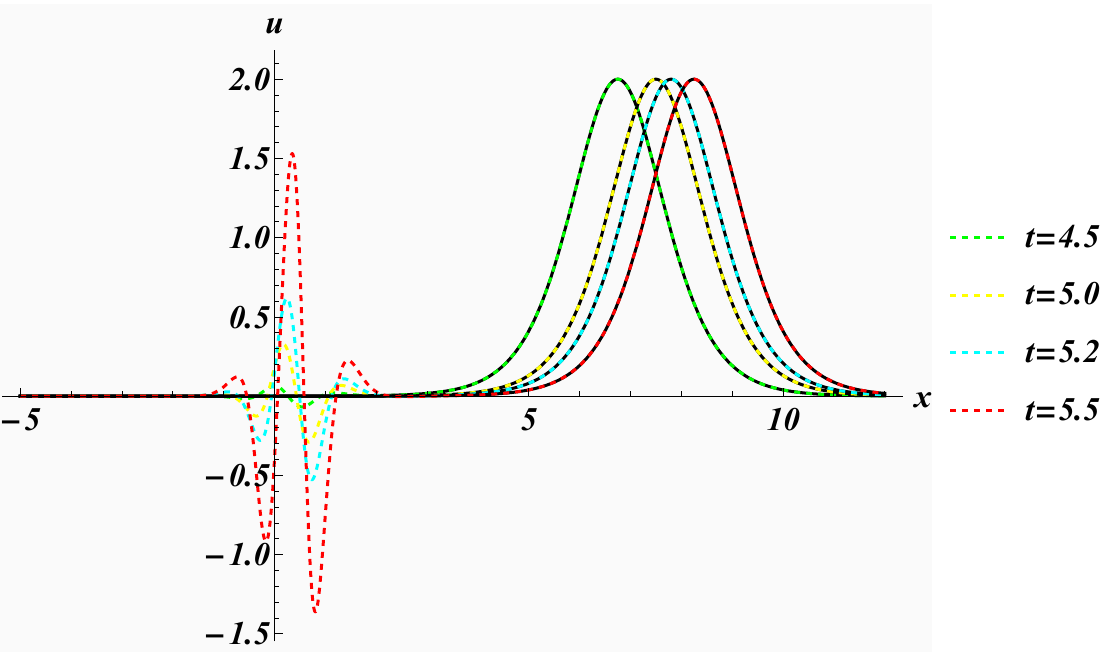}
	\end{minipage}   
	\caption{Evolution of the exact soliton solution (\ref{rotsol}) (solid black) versus the numerical solution (coloured dashed) of the rotated Cauchy problem for the KdV system with initial profiles (\ref{rotexcauch3}) for $\sigma= 3$, $a=2$ together with an emergent chargeless standing wave at the origin. } 
	\label{rotexactkdv}
\end{figure}

As seen in figure \ref{rotexactkdv}, for times up to around $t=4$ the numerical solution smoothly follows the exact solution, but after that a visible singularity starts to develop at the origin in form of an ever growing oscillation which tends to infinity at $t \approx 6.34$. We notice that the oscillations are standing waves that do not make contributions to any of the charges, which for the values used in figure \ref{rotexactkdv} are exactly identical to those obtained from the single soliton solution, i.e. ${\cal Q}_1= 3.26599$, ${\cal Q}_2=2.17732$, ${\cal Q}_3=1.74186$ and ${\cal Q}_4=1.49302$.

\begin{figure}[h]
 	\centering         
 	\begin{minipage}[b]{0.8\textwidth}     
 		\includegraphics[width=\textwidth ]{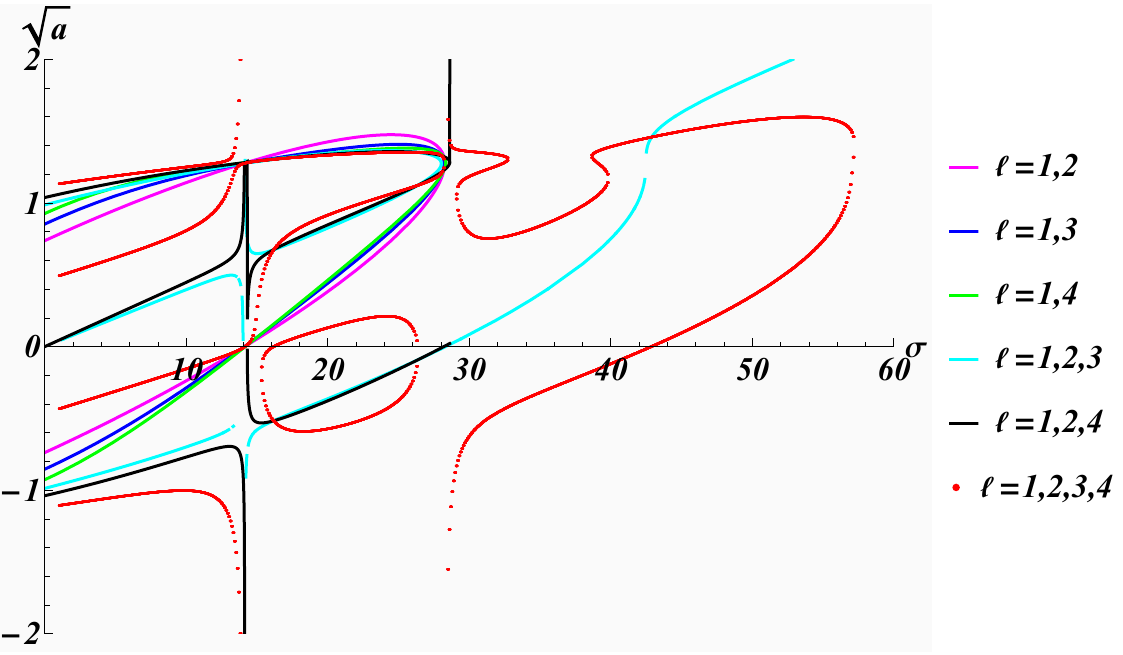}
 	\end{minipage}   
 	\caption{Predicted real square root amplitudes from different combinations of the rotated version of the charge conservation equation (\ref{profevolve}) with Gaussian initial profile in the HDT version of the mKdV-equation. }
 	\label{rotKdVampred}
 \end{figure}

Next we solve the rotated Cauchy problem (\ref{rotstanKdVtr}) for some more generic initial profile, here taken to be a Gaussian as in the previous section. We may then employ the same arguments as for the original unrotated case outlined in section 2.1 and predict the amplitudes of the emerging solitons. It turns out that for the solution (\ref{rotsol}) the values for the charges ${\cal Q}_\ell$ are exactly the same as those computed in (\ref{solcharge}), so that the general expressions for the bounds in (\ref{twosolreg}) remain the same. However, the charges ${\cal Q}_\ell^{(I)}$ for the initial Gaussian profile are different in this case. We find $ {\cal Q}_1^{(I)} = \sqrt{\pi/2^3} $,  $ {\cal Q}_2^{(I)}  = \sqrt{\pi/3^3}$, $ {\cal Q}_3^{(I)}  = \sqrt{\pi/ 4^3}$, $ {\cal Q}_4^{(I)}  = \sqrt{\pi/ 5^3}$. Using the same argument as previously, we find that the two-soliton region (\ref{twosolreg}) is now confined to the interval $\sigma_c < 2 \sigma_c $ with $\sigma_c =16 \times 6^{1/4}/\sqrt{\pi}  \approx  14.1281$. In figure \ref{rotKdVampred} we display the numerical solutions for the real square root amplitudes of the rotated version of the charge conservation equation (\ref{profevolve}).

\begin{figure}[h]
 	\centering         
 	\begin{minipage}[b]{0.49\textwidth}      
 		\includegraphics[width=\textwidth]{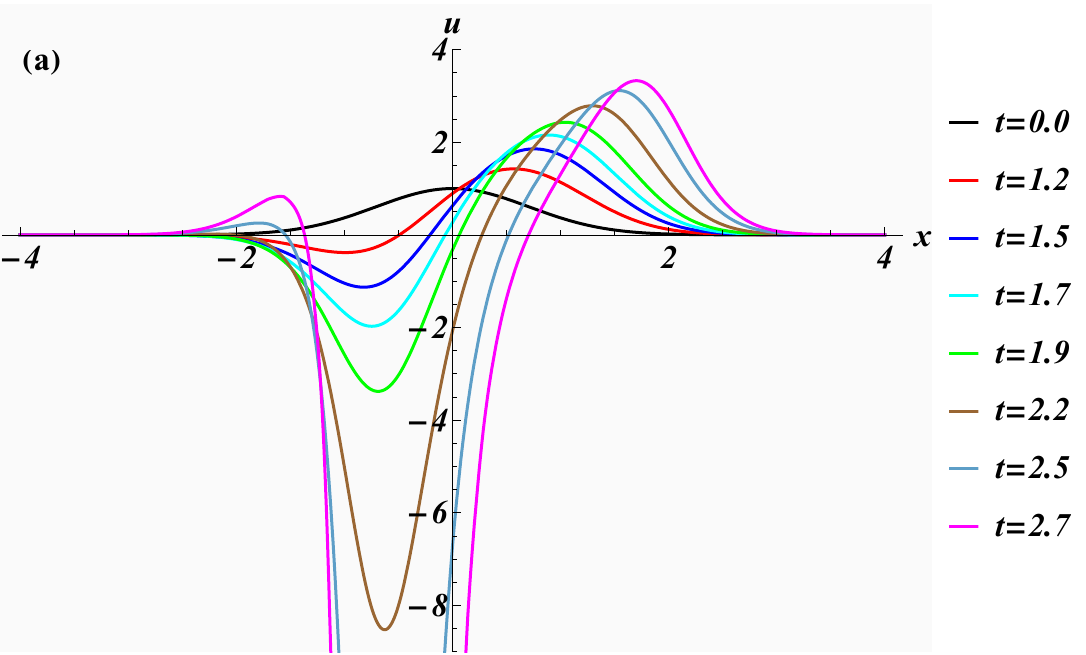}
 	\end{minipage}   
 	\begin{minipage}[b]{0.49\textwidth}      
 		\includegraphics[width=\textwidth]{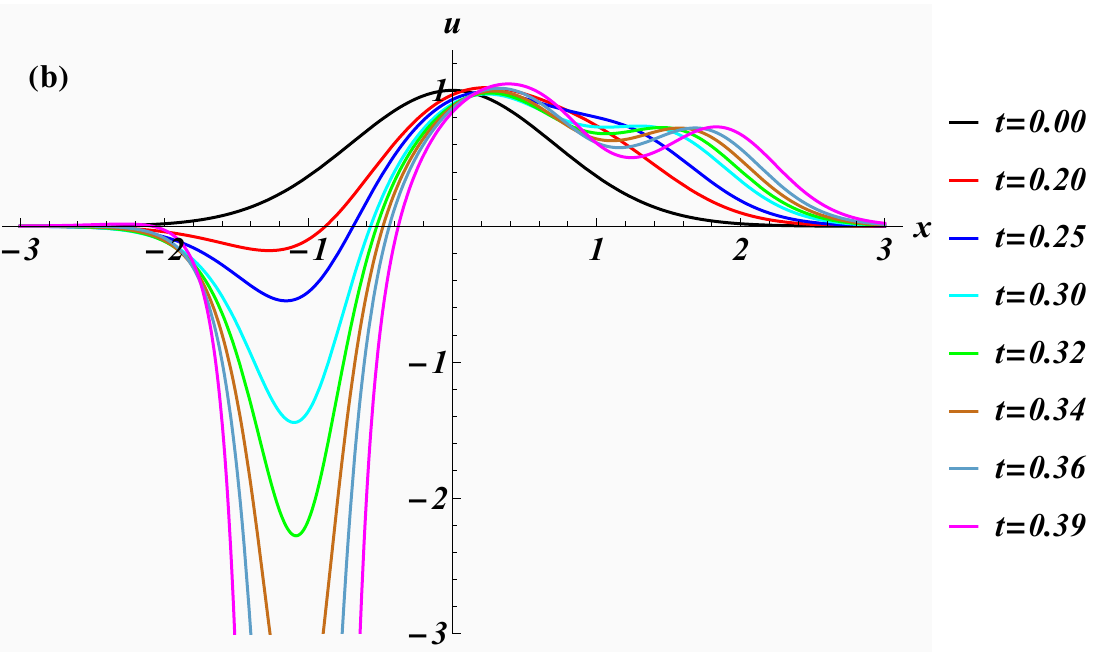}
 	\end{minipage}   
 	\caption{Evolution of a Gaussian, vanishing first and second order time-derivative initial profile for the rotated Cauchy problem of the  KdV equation with $\sigma= 2$ in the nonsoliton region, panel (a) and $\sigma= 27$ in the two-soliton region, panel (b). } 
 	\label{rotgaussiankdv}
 \end{figure}

 Using the requirement that $\sqrt{a_i} \in \mathbb{R}^+$ for $i=1,\ldots, N$, we observe from figure \ref{rotKdVampred} that only in the two soliton region a consistent solution may be found and no $N$-soliton solutions with $N>2$ can be formed. For instance, considering the solution for $\ell=1,2,3$ we observe that in the region $\sigma \lessapprox  39.85 $ always one of the solutions is negative, whereas for $\sigma \gtrapprox  39.85 $ only one of the solutions is real. Hence, no consistent three-soliton solution can be found. Indeed, this feature is confirmed by our numerical solutions shown in figure \ref{rotgaussiankdv}, for the initial profile $u(x,0)=e^{-x^2}$, $u_t(x,0)=u_{tt}(x,0)=0$ and vanishing asymptotic conditions in $x$. 
 
 At the same time these predictions combine with the emerging of a singularity. For values of $\sigma $ in the nonsoliton region $\sigma < \sigma_c$  we observe the emergence of a ``defected'' one-soliton and a single peakon that evolves into a single peak singularity. In contrast, for the larger values of $ \sigma$ in the two-soliton region $ \sigma_c <\sigma  <2 \sigma_c$, as predicted by (\ref{twosolreg}), we see that the wave indeed starts to morphe into a two one-soliton structure, but before they are fully developed, the singularity in time has already occurred. We have verified that while the profiles evolve, the charges are conserved remaining ${\cal Q}_1= 0.627$, ${\cal Q}_2=0.341$, ${\cal Q}_3=0.222$, ${\cal Q}_4=0.159$ in both cases. For various values of $\sigma > 2\sigma_c$ we have also verified that no $N$-soliton, not even in some indicated infant stage, begins to emerge. This agrees precisely with our predictions resulting from the charge conservation equations.

\subsection{Emergent solitons in the HTDT version of the modified KdV system}
\begin{figure}[h]
	\centering         
	\begin{minipage}[b]{0.32\textwidth}      
		\includegraphics[width=\textwidth]{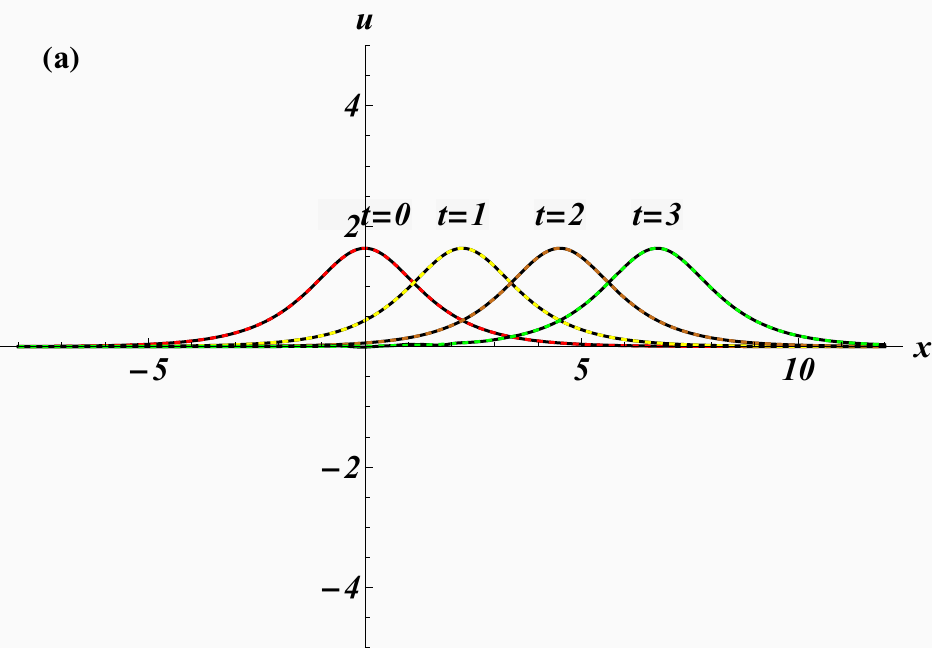}
	\end{minipage}   
	\begin{minipage}[b]{0.32\textwidth}      
		\includegraphics[width=\textwidth]{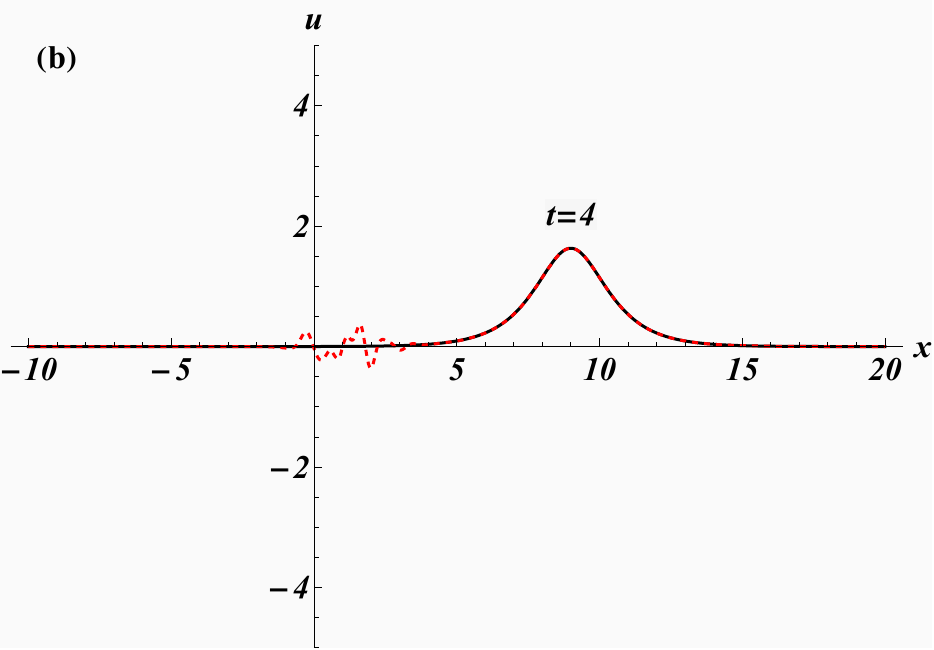}
	\end{minipage}   
	\begin{minipage}[b]{0.32\textwidth}      
		\includegraphics[width=\textwidth]{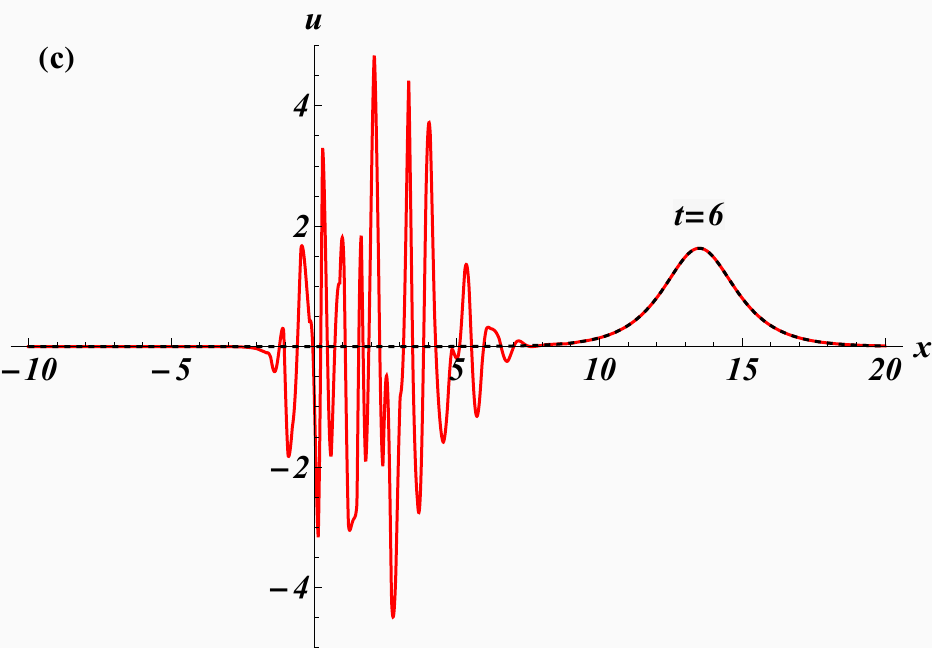}
	\end{minipage}   
	\begin{minipage}[b]{0.32\textwidth}      
		\includegraphics[width=\textwidth]{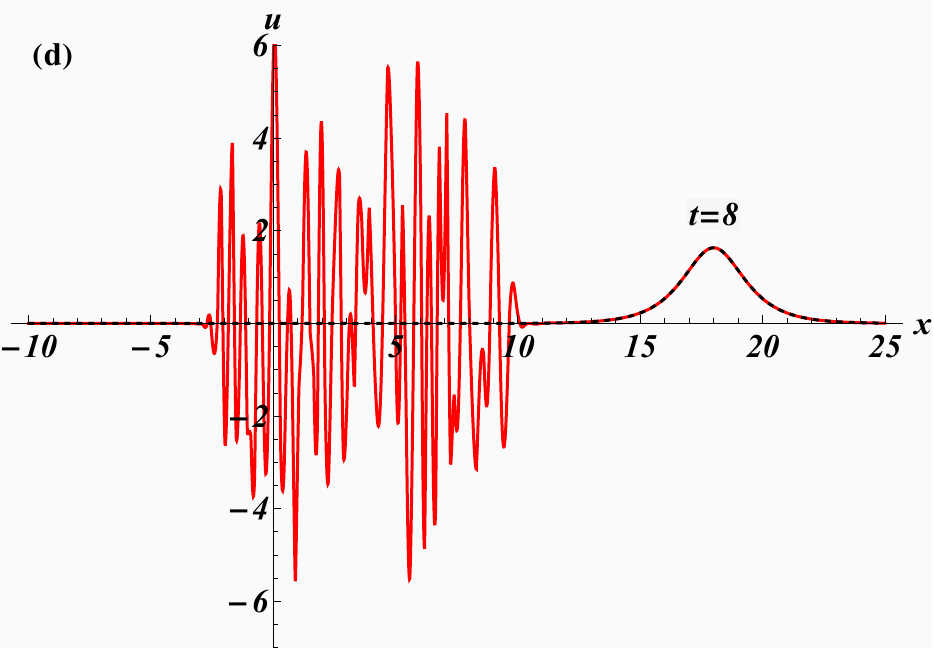}
	\end{minipage}   
	\begin{minipage}[b]{0.32\textwidth}      
		\includegraphics[width=\textwidth]{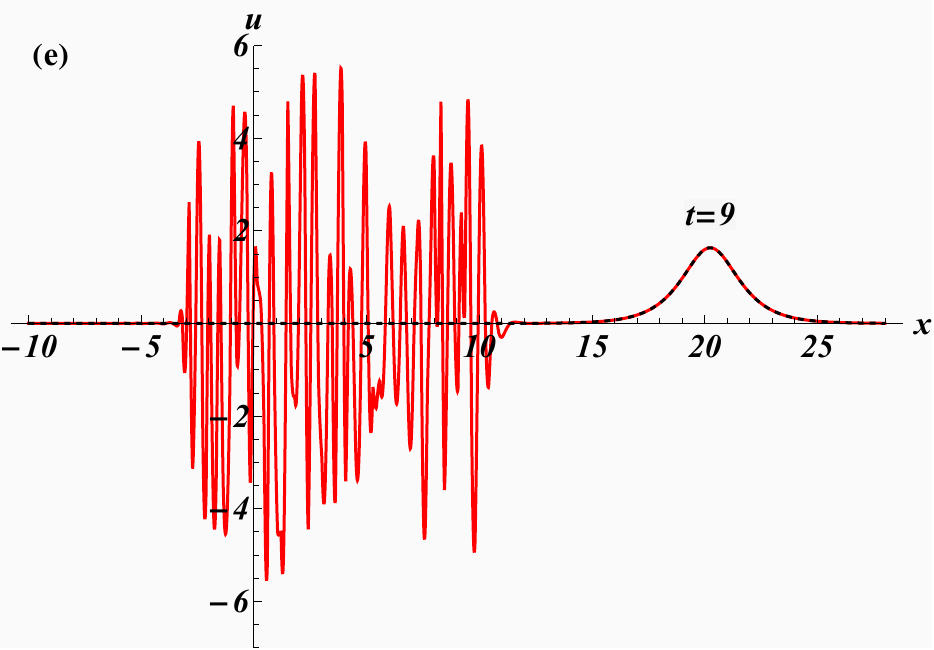}
	\end{minipage}   
	\begin{minipage}[b]{0.32\textwidth}      
		\includegraphics[width=\textwidth]{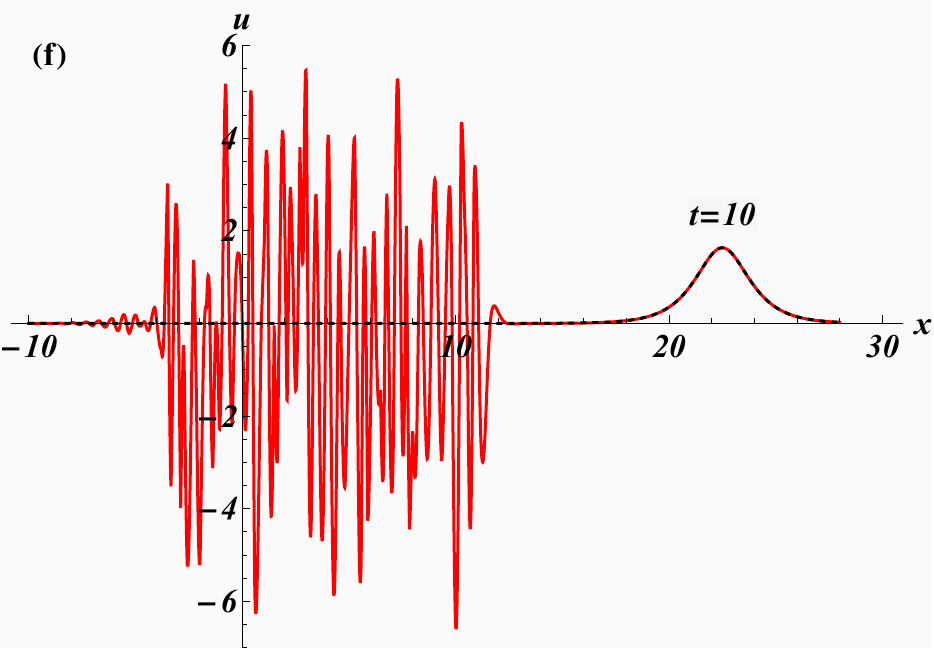}
	\end{minipage}   
	\caption{Evolution of the exact soliton solution (\ref{rotsolmkdv}) (dashed black) versus the numerical solution (red solid) of the rotated Cauchy problem for the modified KdV system with initial profiles (\ref{rotexcauch}) with an emergent standing wave at the origin for $\sigma= 3$ and $a =2 \sqrt{2/3}$.  } 
	\label{rotmkdvexact}
\end{figure}

For the case $n=4$ in (\ref{rotstanKdVtr}) we find the exact one-soliton solution
\begin{equation}
	u(x,t) =   a \, \text{sech}\left[  \frac{a^3 \sigma }{6 \sqrt{6}} \left( x -  \frac{6}{a^2}   t      \right)  \right]   . \label{rotsolmkdv}
\end{equation} 
At first we track this exact solution with the initial profiles directly corresponding to (\ref{rotsolmkdv})
\begin{eqnarray}
	u(x,0)&=& a \, \text{sech}\left( \frac{a^3 \sigma }{6 \sqrt{6}}  x  \right)  , \qquad \,\,\,  u_t(x,0)= \frac{a^2 \sigma  }{\sqrt{6}}  \tanh \left(\frac{a^3 \sigma  x}{6 \sqrt{6}}\right) \text{sech}\left(\frac{a^3 \sigma  x}{6
		\sqrt{6}}\right)  , \,\,\,\,\,  \label{rotexcauch} \\
	u_{tt}(x,0)&=& \frac{a^3 \sigma ^2 }{12} \left[\cosh \left(\frac{a^3 \sigma  x}{3 \sqrt{6}}\right)-3\right]
	\text{sech}^3\left(\frac{a^3 \sigma  x}{6 \sqrt{6}}\right)  , \,\,\, \quad \lim_{\vert x \vert \rightarrow \infty}u( x,t) = 0. \notag
\end{eqnarray}
Unlike as in the HTD version of the KdV systems we can track this exact one-soliton solution quite precisely to arbitrary large time as depicted in figure \ref{rotmkdvexact}. 

We also observe that similarly as for the HTD version of the KdV system oscillations start to emerge near the origin, but crucially in this case they remain finite in amplitude. 

\begin{figure}[h]
	\centering         
	\begin{minipage}[b]{0.8\textwidth}     
		\includegraphics[width=\textwidth ]{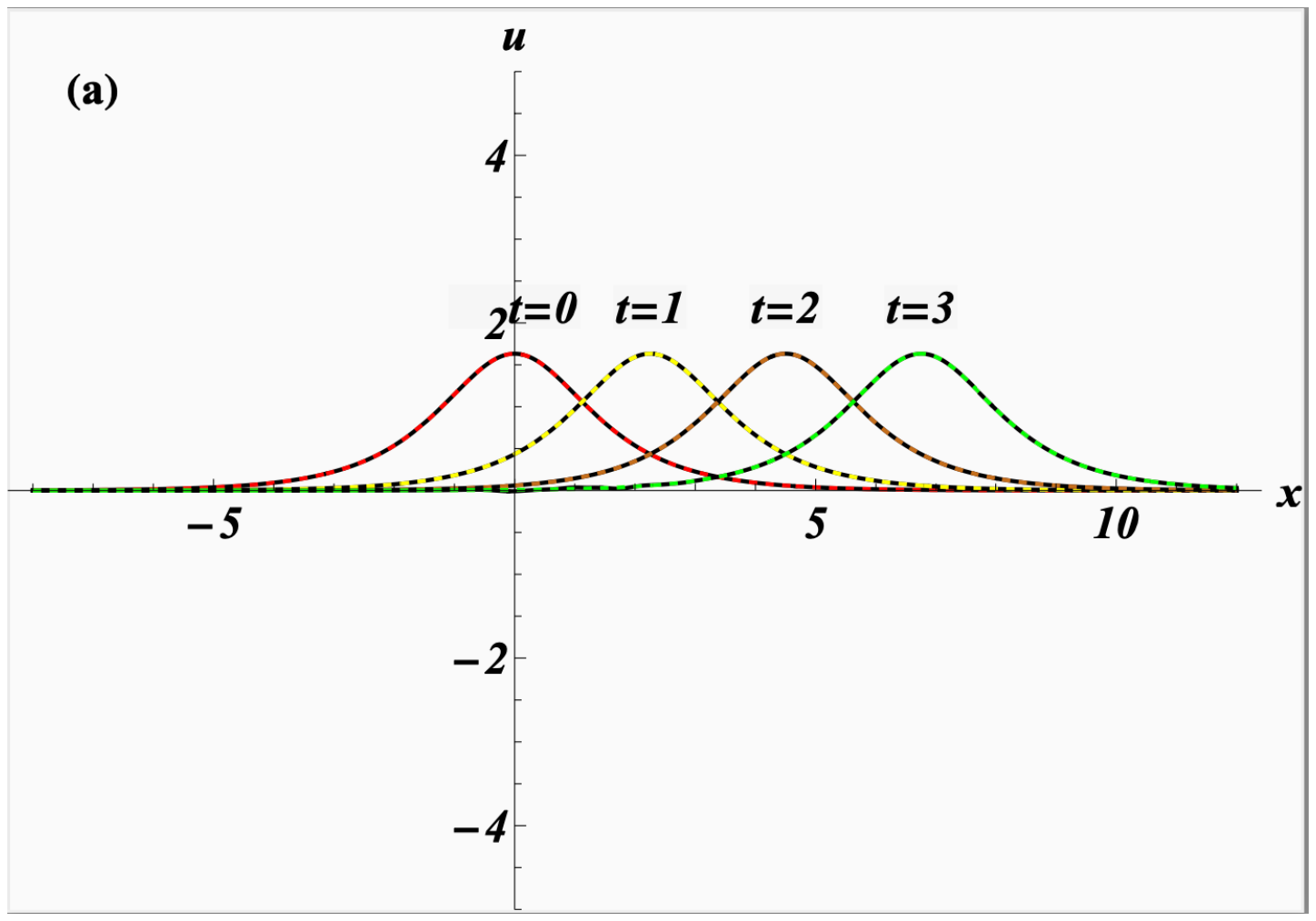}
	\end{minipage}   
	\caption{Predicted real amplitudes from different combinations of the rotated version of the charge conservation equation (\ref{profevolve}) with Gaussian initial profile in the HDT version  of the mKdV-equation.}
	\label{rotmKdVampred}
\end{figure}

Next we probe what happens when we send in an arbitrary initial profile. The charges corresponding to the solution (\ref{rotsolmkdv}) are the same as in the unrotated case (\ref{concharmKdV}). However, the charges corresponding to the Gaussian initial profile are different
\begin{equation}
	{\cal Q}_n^{(I)} = \sqrt{ \frac{\pi  }{ 8 \left( n+1\right)^3  } } , \qquad  n=1,2,\ldots
\end{equation} 
so that the predictions for the amplitudes from the conservation laws will also vary. Our predictions are depicted in figure \ref{rotmKdVampred}.

Unlike as in the HTD version of KdV, now negative amplitudes in the solution (\ref{rotsolmkdv}) are permitted, since $a \rightarrow - a$ leads to $u \rightarrow - u$, which is also a solution of (\ref{rotstanKdVtr}) for $n$ even. Thus we see that for $\sigma \lessapprox  39.85 $ all of the calculated possibilities for $N$-soliton solutions are actually realised, i.e., $N=2,3,4,5$. We conjecture that this will hold also beyond the cases we have computed for all $N >5$. In the region for  $\sigma \gtrapprox  39.85 $ only solutions for $N >4$ are acquired. 

\begin{figure}[h]
	\centering         
	\begin{minipage}[b]{0.32\textwidth}      
		\includegraphics[width=\textwidth]{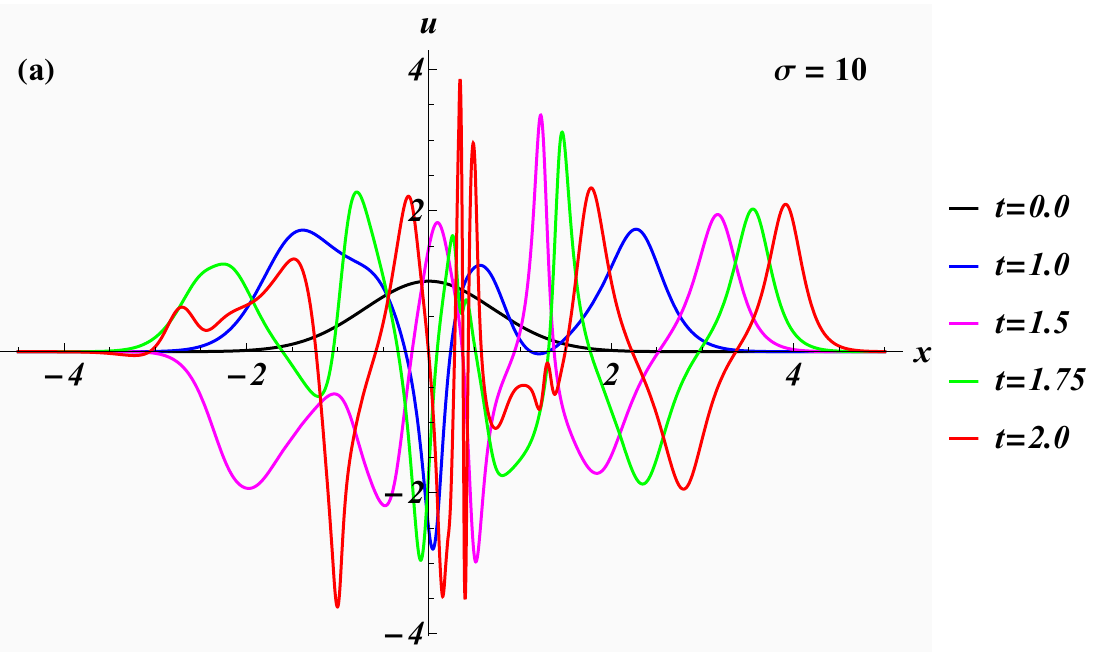}
	\end{minipage}   
	\begin{minipage}[b]{0.32\textwidth}      
		\includegraphics[width=\textwidth]{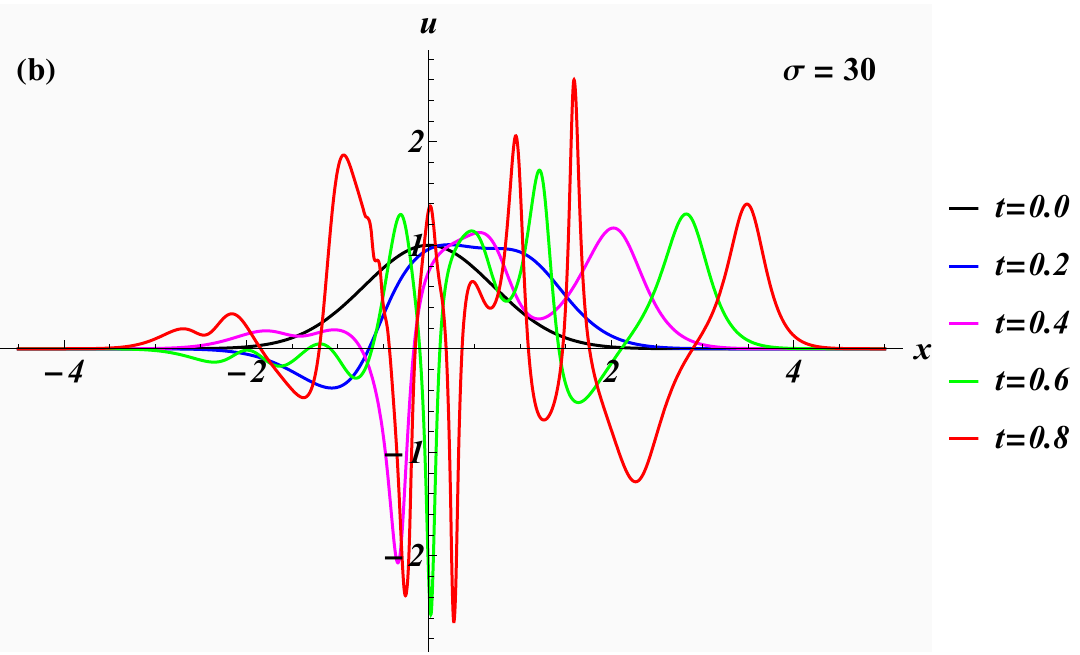}
	\end{minipage}   
	\begin{minipage}[b]{0.32\textwidth}      
		\includegraphics[width=\textwidth]{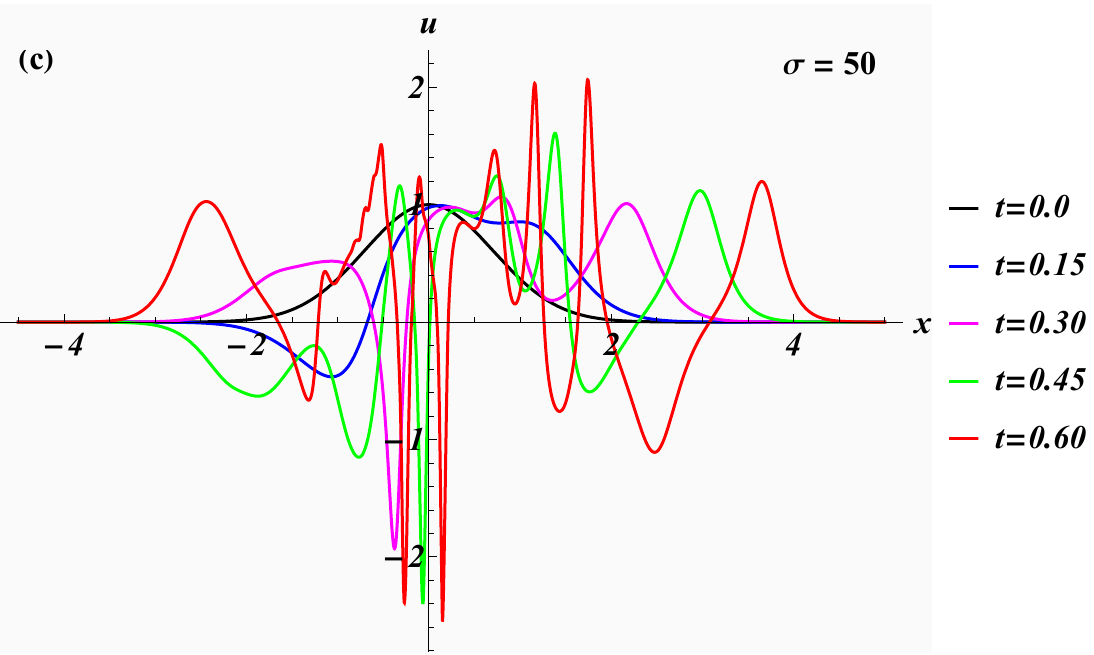}
	\end{minipage}     
	\caption{Evolution of a Gaussian, vanishing first and second order time-derivative initial profile for the HTD version of the modified KdV equation in different characteristic regions. } 
	\label{rotmkdvgausss}
\end{figure}

The overall effect on the evolution of the initial profile is that the localised wave tries to decay into $N$-soliton soltions with larger and large $N$ as time evolves. We conjecture that this feature is responsible for the oscillatory behaviour as seen in figure \ref{rotmkdvgausss}. In the positive $x$-region we can identify the various $N$-soliton solutions that can be realised at different times and notice further that for larger values of $\sigma$ the larger $N$-soliton solutions are settled into much quicker. At the same time the overall behaviour remains {\em benign} (or {\em metastable} \cite{Salvio1}), in the sense that all the solutions for the amplitudes predicted from different conservation laws are finite.

\subsection{Solitary waves in nonintegrable HTD mKdV systems}

Finally, we investigate the HTD systems with $n>4$, which are not integrable. The latter means that we do not expect soliton solutions to appear, but also that all the restrictions imposed by the conservation laws are absent. Our explicit computations show that all models with $n$ odd and $n>4$ behave for all $\sigma$ qualitatively in the same manner as the $n=3$ theory in the nonsoliton regime. For $n$ even and $n>4$ we observe a similar, but more random behaviour as in the $n=4$ model.

\section{Conclusions}

In the first part we have revisited the problem of how an initial localised profile evolves when propagated by means of nonlinear modified KdV systems. Exploiting the integrability of some of these systems, we used various combinations of conservation laws to predict the number of solitons into which the profile will be permitted to settle into, as well as their respective amplitudes. By refining the previously carried out analysis, we found that an initial profile will always decay into the maximal number of $N$-solitons that is allowed by the conservation laws. We conjecture that this is a general feature. For the nonintegrable versions of these theories we found that in the $n=5$ case the features of the nonsoliton regime of the integrable systems are still present. In that case a solitary wave moving at constant speed emerged together with an oscillatory tail at negative $x$. No decay into multi-solitons was observed. For all other cases with we found that the profile will always evolve into the oscillatory tail that will eventually distribute the charge into all modes.

When adapting the analysis to the higher time-derivative versions of these theories, we derived that the only allowed breakup, by integrability, of the initial profile is into a two-soliton solution for the $n=3$ case. This feature then combines with the previously observed property that the classical solutions of these theories will develop instabilities \cite{Smilga6,Smilgaacta,fring2024higher}, as is to be expected in HTDT. For the exact solutions of the HTD-KdV system the singularities manifest themselves as chargeless standing waves at the origin, whose amplitudes grow to infinity as time evolves. Instead, in the HTD-mKdV system we found that the profile is allowed to settle into any of the $N$-soliton solutions, which gives rise to the oscillations spreading out from the origin.
As the solutions for all of the predicted amplitudes is finite, these oscillations do not grow 
to infinity. For the nonintegrable theories with $n$ odd we found the same behaviour as for the $n=3$ theory in the nonsoliton region. For the cases with $n$ even we found that the disturbance settles into a more random set of oscillations.  We noticed that the soliton/solitary waves in the HTDT are usually slower when compared to their first order time-derivative counterparts. The spreading speed of the oscillations is faster in all observed cases so that the solitary wave structures were always found to be absorbed by the oscillations spreading out.  

There are some obvious open questions. Here we have always taken a simple Gaussian as initial profile $u(x,0)=e^{-x^2}$, and have set the independent profiles for the first and second order time-derivative to zero. We found that changing the time derivative profiles does not change the overall characteristic behaviour, but a more systematic analysis, using different options for these profiles, would be interesting to obtain. It would be especially insightful to find out whether it is possible to prolong the lifetime of the soliton/solitary wave structures to such an extend that they can fully develop before being absorbed by the oscillations. Evidently, it would be interesting to develop analytical arguments that predict the speed of the spread of the oscillation study and to study the observed effect in other types of integrable models.

\medskip
\noindent \textbf{Acknowledgments:}  TT is supported by JSPS KAKENHI Grant Number 22KJ0752. BT is supported by a City, University of London Research Fellowship.  AF would like to thank Artur Sergyeyev for bringing reference 37 to our attention.

\newif\ifabfull\abfulltrue


\begin{thebibliography}{10}
	
	\bibitem{zabusky1965int}
	N.~Zabusky and M.~D. Kruskal,
	\newblock Interaction of ``solitons" in a collisionless plasma and the
	recurrence of initial states,
	\newblock Phys. Rev. Lett. {\bf 15}(6), 240 (1965).
	
	\bibitem{fermi1955studies}
	E.~Fermi, P.~Pasta, S.~Ulam, and M.~Tsingou,
	\newblock Studies of the nonlinear problems,
	\newblock Technical report, Los Alamos National Lab.(LANL), Los Alamos, NM
	(United States), 1955.
	
	\bibitem{berman2005fermi}
	G.~P. Berman and F.~M. Izrailev,
	\newblock The Fermi--Pasta--Ulam problem: fifty years of progress,
	\newblock Chaos {\bf 15}(1) (2005).
	
	\bibitem{berezin}
	Y.~A. Berezin and V.~I. Karpman,
	\newblock Nonlinear evolution of disturbances in plasmas and other dispersive
	media,
	\newblock Sov. Phys. JETP {\bf 24}(5), 1049--1056 (1967).
	
	\bibitem{jeffrey1972weak}
	A.~Jeffrey and T.~Kakutani,
	\newblock Weak nonlinear dispersive waves: a discussion centered around the
	Korteweg--de Vries equation,
	\newblock Siam Review {\bf 14}(4), 582--643 (1972).
	
	\bibitem{pais1950field}
	A.~Pais and G.~E. Uhlenbeck,
	\newblock On field theories with non-localized action,
	\newblock Phys. Rev. {\bf 79}(1), 145 (1950).
	
	\bibitem{stelle77ren}
	K.~S. Stelle,
	\newblock Renormalization of higher-derivative quantum gravity,
	\newblock Phys. Rev. D {\bf 16}(4), 953 (1977).
	
	\bibitem{grav1}
	A.~A. Starobinsky,
	\newblock A new type of isotropic cosmological models without singularity,
	\newblock Phys. Lett. B {\bf 91}(1), 99--102 (1980).
	
	\bibitem{grav2}
	S.~L. Adler,
	\newblock Einstein gravity as a symmetry-breaking effect in quantum field
	theory,
	\newblock Rev. Mod. Phys. {\bf 54}(3), 729 (1982).
	
	\bibitem{grav3}
	A.~V. Smilga,
	\newblock Spontaneous generation of the Newton constant in the renormalizable
	gravity theory,
	\newblock ITEP preprint 63 (1982) 8 pp, arXiv preprint arXiv:1406.5613 (2014)
	(1982).
	
	\bibitem{modesto16super}
	L.~Modesto and I.~L. Shapiro,
	\newblock Superrenormalizable quantum gravity with complex ghosts,
	\newblock Phys. Lett. B {\bf 755}, 279--284 (2016).
	
	\bibitem{ghostconst}
	T.-J. Chen, M.~Fasiello, E.~A. Lim, and A.~J. Tolley,
	\newblock Higher derivative theories with constraints: Exorcising
	Ostrogradski's Ghost,
	\newblock J. Cos. Astro. Phys. {\bf 2013}(02), 042 (2013).
	
	\bibitem{salvio16quant}
	A.~Salvio and A.~Strumia,
	\newblock Quantum mechanics of 4-derivative theories,
	\newblock The EPJ C {\bf 76}, 1--15 (2016).
	
	\bibitem{fakeons}
	D.~Anselmi,
	\newblock Fakeons and Lee-Wick models,
	\newblock JHEP {\bf 2018}(2) (2018).
	
	\bibitem{bender2008no}
	C.~M. Bender and P.~D. Mannheim,
	\newblock No-ghost theorem for the fourth-order derivative Pais-Uhlenbeck
	oscillator model,
	\newblock Phys. Rev. Lett. {\bf 100}(11), 110402 (2008).
	
	\bibitem{raidal2017quantisation}
	M.~Raidal and H.~Veerm{\"a}e,
	\newblock On the quantisation of complex higher derivative theories and
	avoiding the Ostrogradsky ghost,
	\newblock Nucl. Phys. B {\bf 916}, 607--626 (2017).
	
	\bibitem{Hawking}
	S.~W. Hawking and T.~Hertog,
	\newblock Living with ghosts,
	\newblock Phys. Rev. D {\bf 65}(10), 103515 (2002).
	
	\bibitem{biswas2010towards}
	T.~Biswas, T.~Koivisto, and A.~Mazumdar,
	\newblock Towards a resolution of the cosmological singularity in non-local
	higher derivative theories of gravity,
	\newblock JCAP {\bf 2010}(11), 008 (2010).
	
	\bibitem{Salvio2}
	A.~Salvio,
	\newblock Dimensional transmutation in gravity and cosmology
	\newblock Int. J. Mod. Phys. A {\bf 36}, 2130006 (2021).
	
	\bibitem{Salvio3}
	A.~Salvio,
	\newblock Quasi-conformal models and the early universe
	\newblock Europ. Phys. J. C {\bf 79}, 750 (2019).
	
	\bibitem{Salvio4}
	A.~Salvio,
	\newblock A non-Perturbative and Background-Independent Formulation of Quadratic Gravity
	\newblock arXiv preprint arXiv:2404.08034.
	
	\bibitem{weldon98finite}
	H.~A. Weldon,
	\newblock Finite-temperature retarded and advanced fields,
	\newblock Nucl. Phys. B {\bf 534}(1-2), 467--490 (1998).
	
	\bibitem{mignemi1992black}
	S.~Mignemi and D.~L. Wiltshire,
	\newblock Black holes in higher-derivative gravity theories,
	\newblock Phys. Rev. D {\bf 46}(4), 1475 (1992).
	
	\bibitem{rivelles2003triviality}
	V.~O. Rivelles,
	\newblock Triviality of higher derivative theories,
	\newblock Phys. Lett. B {\bf 577}(3-4), 137--142 (2003).
	
	\bibitem{Kap1}
	D.~S. Kaparulin, S.~L. Lyakhovich, and A.~A. Sharapov,
	\newblock BRST analysis of general mechanical systems,
	\newblock J. of Geo. and Phys. {\bf 74}, 164--184 (2013).
	
	\bibitem{plyush89mass}
	M.~S. Plyushchay,
	\newblock Massless point particle with rigidity,
	\newblock Mod. Phys. Lett. A {\bf 4}(09), 837--847 (1989).
	
	\bibitem{Mpl}
	M.~S. Plyushchay,
	\newblock Massless particle with rigidity as a model for the description of
	bosons and fermions,
	\newblock Phys. Lett. B {\bf 243}(4), 383--388 (1990).
	
	\bibitem{dine1997comments}
	M.~Dine and N.~Seiberg,
	\newblock Comments on higher derivative operators in some SUSY field theories,
	\newblock Phys. Lett. B {\bf 409}(1-4), 239--244 (1997).
	
	\bibitem{smilga17ultrav}
	A.~Smilga,
	\newblock Ultraviolet divergences in non-renormalizable supersymmetric
	theories,
	\newblock Phys. of Part. and Nucl. Lett. {\bf 14}, 245--260 (2017).
	
	\bibitem{Sugg1}
	M.~Pav{\v{s}}i{\v{c}},
	\newblock Stable self-interacting Pais--Uhlenbeck oscillator,
	\newblock Mod. Phys. Lett. A {\bf 28}(36), 1350165 (2013).
	
	\bibitem{Sugg2}
	D.~S. Kaparulin, S.~L. Lyakhovich, and A.~A. Sharapov,
	\newblock Classical and quantum stability of higher-derivative dynamics,
	\newblock EPJ C {\bf 74}, 1--19 (2014).
	
	\bibitem{Sugg3}
	M.~Avendano-Camacho, J.~A. Vallejo, and Y.~Vorobiev,
	\newblock A perturbation theory approach to the stability of the Pais-Uhlenbeck
	oscillator,
	\newblock J. Math. Phys. {\bf 58}(9) (2017).
	
	\bibitem{Sugg4}
	N.~Boulanger, F.~Buisseret, F.~Dierick, and O.~White,
	\newblock Higher-derivative harmonic oscillators: stability of classical
	dynamics and adiabatic invariants,
	\newblock EPJ C {\bf 79}, 1--8 (2019).
	
	\bibitem{deffayet22ghost}
	C.~Deffayet, S.~Mukohyama, and A.~Vikman,
	\newblock Ghosts without runaway instabilities,
	\newblock Phys. Rev. Lett. {\bf 128}(4), 041301 (2022).
	
	\bibitem{deffayet23global}
	C.~Deffayet, A.~Held, S.~Mukohyama, and A.~Vikman,
	\newblock Global and local stability for ghosts coupled to positive energy
	degrees of freedom,
	\newblock JCAP {\bf 2023}(11), 031 (2023).
	
	\bibitem{smilga2021exactly}
	A.~Smilga,
	\newblock On exactly solvable ghost-ridden systems,
	\newblock Phys. Lett. A {\bf 389}, 127104 (2021).
	
	\bibitem{Samoilenko}
	V.~G. Samoilenko and N.~N. Pritula and U.~S. Suyarov, 
	\newblock  The complete integrability analysis of the inverse Korteweg-de Vries equation
	\newblock Ukrainian Math. J. {\bf 43}, 1157--1164 (1991).
	
	\bibitem{bethanAF}
	A.~Fring and B.~Turner,
	\newblock Higher derivative Hamiltonians with benign ghosts from affine Toda
	lattices,
	\newblock J. Phys. A: Math. Theor. {\bf 56}, 295203 (2023).
	
	\bibitem{fring23int}
	A.~Fring and B.~Turner,
	\newblock Integrable scattering theory with higher derivative Hamiltonians,
	\newblock Eur. Phys. J. Plus {\bf 138}(12), 1136 (2023).
	
	\bibitem{fring2024higher}
	A.~Fring, T.~Taira, and B.~Turner,
	\newblock Higher Time-Derivative Theories from Space--Time Interchanged
	Integrable Field Theories,
	\newblock Universe {\bf 10}(5), 198 (2024).
	
	\bibitem{Smilga6}
	T.~Damour and A.~Smilga,
	\newblock Dynamical systems with benign ghosts,
	\newblock Phys. Rev. D {\bf 105}(4), 045018 (2022).
	
	\bibitem{Smilgaacta}
	A.~Smilga,
	\newblock Modified Korteweg-de Vries equation as a system with benign ghosts,
	\newblock Acta Pol. {\bf 61}(1), 190--196 (2022).
	
	\bibitem{whitham1965non}
	G.~B. Whitham,
	\newblock Non-linear dispersive waves,
	\newblock Proc. R. Soc. Lond. A. Math. and Phys. Sci. {\bf 283}(1393), 238--261
	(1965).
	
	\bibitem{miura1968kortxx}
	R.~M. Miura,
	\newblock Korteweg-de Vries equation and generalizations. I. A remarkable
	explicit nonlinear transformation,
	\newblock J. Math. Phys. {\bf 9}(8), 1202--1204 (1968).
	
	\bibitem{miura1968korteweg}
	R.~M. Miura, C.~S. Gardner, and M.~D. Kruskal,
	\newblock Korteweg-de Vries equation and generalizations. II. Existence of
	conservation laws and constants of motion,
	\newblock J. Math. Phys. {\bf 9}(8), 1204--1209 (1968).
	
	\bibitem{Nutku}
	Y.~Nutku,
	\newblock Hamiltonian formulation of the KdV equation,
	\newblock J. Math. Phys. {\bf 25}, 2007--2008 (1984).
	
	\bibitem{karpman67asymp}
	V.~I. Karpman,
	\newblock An asymptotic solution of the Korteweg-de Vries equation,
	\newblock Phys. Lett. A {\bf 25}(10), 708--709 (1967).
	
	\bibitem{Salvio1}
	A.~Salvio,
	\newblock Metastability in quadratic gravity
	\newblock Phys. Rev. D {\bf 99}, 103507 (2019).
	
	
\end{thebibliography}

\end{document}